\documentclass[a4paper,onecolumn,11pt]{quantumarticle}
\pdfoutput=1
\usepackage[utf8]{inputenc}
\usepackage[english]{babel}
\usepackage[T1]{fontenc}
\usepackage{amsmath}
\usepackage{hyperref}
\usepackage[numbers,sort&compress]{natbib} 

\usepackage{tikz}
\usepackage{lipsum}
\usepackage{booktabs}
\usepackage{diagbox} 
\usepackage{multirow} 
\usepackage{makecell}
\usepackage{longtable}
\usepackage{array}
\usepackage{tabularx}
\usepackage{algorithmic}
\usepackage{algorithm}
\usepackage{graphicx}    
\usepackage{qcircuit}
\usepackage{xcolor}  
\usepackage{tikz}  
\usepackage{geometry}
\usepackage{graphicx} 
\usepackage{subfigure}
\usepackage{arydshln}
\usepackage{bm}
\usepackage[percent]{overpic}
\usepackage{threeparttable}
\usepackage{bbm}
\usepackage{indentfirst}
\usetikzlibrary{arrows, shapes, chains}

\begin{document}

\title{Classical-Assisted Quantum Ground State Preparation with Tensor Network States and Monte Carlo Sampling}

\author[1,2]{Feng-Yu Le}
\author[2]{Zhao-Yun Chen${^*}$} \email{chenzhaoyun@iai.ustc.edu.cn}
\author[1,2]{Lu Wang}
\author[2]{Cheng Xue}
\author[2]{Chao Wang}
\author[2,3,4,5]{Yong-Jian Han}
\author[2,3,4,5]{Yu-Chun Wu}
\author[7]{Qing Yan}
\author[2]{Shaojun Dong$^{*}$}
\email{sj.dong@iai.ustc.edu.cn}
\author[2,3,4,5,6]{Guo-Ping Guo}
\affil[1]{AHU-IAI AI Joint Laboratory, Anhui University, Hefei, Anhui, 230601, P. R. China}
\affil[2]{Institute of Artificial Intelligence, Hefei Comprehensive National Science Center, Hefei, Anhui, 230088, P. R. China}
\affil[3]{CAS Key Laboratory of Quantum Information, University of Science and Technology of China, Hefei, Anhui, 230026, P. R. China}
\affil[4]{CAS Center For Excellence in Quantum Information and Quantum Physics, University of Science and Technology of China, Hefei, Anhui, 230026, P. R. China}
\affil[5]{Hefei National Laboratory, Hefei, Anhui, 230088, P. R. China}
\affil[6]{Origin Quantum Computing Company Limited, Hefei, Anhui, 230088, P. R. China}
\affil[7]{College of Electrical Engineering and Automation, Anhui University, Hefei, Anhui 230601, P. R. China}

\begin{abstract}
   Quantum computing offers potential solutions for finding ground states in condensed-matter physics and chemistry. However, achieving effective ground state preparation is also computationally hard for arbitrary Hamiltonians. It is necessary to propose certain assumptions to make this problem efficiently solvable, including preparing a trial state of a non-trivial overlap with the genuine ground state. Here, we propose a classical-assisted quantum ground state preparation method for quantum many-body systems, combining Tensor Network States (TNS) and Monte Carlo (MC) sampling as a heuristic method to prepare a trial state with a non-trivial overlap with the genuine ground state. We extract a sparse trial state by sampling from TNS, which can be efficiently prepared by a quantum algorithm on early fault-tolerant quantum computers. Our method demonstrates a polynomial improvement in scaling of overlap between the trial state and genuine ground state compared to random trial states, as evidenced by numerical tests on the spin-$1/2$ $J_1$-$J_2$ Heisenberg model. Furthermore, our method is a novel approach to hybridize a classical numerical method and a quantum algorithm and brings inspiration to ground state preparation in other fields.
\end{abstract}

\section{Introduction}
\label{section1}
Quantum simulation plays a crucial role in many realms including condensed-matter physics, high-energy physics, atomic physics, and quantum chemistry~\cite{georgescu2014quantum}. One important aspect is the ground state preparation of quantum many-body interacting systems, which describes the most interesting phases of matter, such as quantum magnetism and high-temperature superconductivity~\cite{vasiliev2018milestones,buzzi2020photomolecular,broholm2020quantum,balents2010spin}.
However, the computational cost of accurately calculating the ground state scales exponentially with the size of the system~\cite{kandala2017hardware}. This \emph{exponential explosion} problem has hindered the development of classical efficient methods for quantum many-body systems~\cite{luitz2015many,liu2017gradient}.
Quantum computing has emerged as an alternative approach for exploring the behavior and properties of ground states that may provide solutions and potentially outperform classical methods in terms of efficiency~\cite{huggins2022unbiasing}.

Most existing quantum ground state preparation algorithms concentrate on Hamiltonian-agnostic configuration which is hopefully applied to random Hamiltonian. 
However, efficient ground state preparation cannot be feasible for all Hamiltonians, since the task of local Hamiltonians is proven to be QMA-complete~\cite{poulin2009preparing,lemieux2021resource}. Many algorithms propose certain assumptions to make this problem efficiently solvable~\cite{farhi2014quantum,huggins2020non,ge2019faster,poulin2009preparing,lemieux2021resource,lin2020near,lin2022heisenberg}.
It is noteworthy that quantum algorithms with provable performance guarantees require the preparation of a trial state with a non-trivial overlap with the genuine ground state as a prerequisite~\cite{lin2020near,lin2022heisenberg}.
Moreover, the time complexity of these algorithms is affected by the overlap, which tends to decrease exponentially with the size of the system, especially employing random trial states.
Hence, improving the overlap holds great potential for significantly improving the performance of quantum ground state preparation methods.

Substantial progress has been made in preparing trial states with a non-trivial overlap for quantum chemistry application.
Adiabatic algorithms~\cite{whitfield2011simulation}, Hartree-Fock state, and Variational Quantum Eigensolver (VQE)~\cite{halder2021digital} method are proposed to be a heuristic method to first obtain a trial state~\cite{oh2008quantum}, which works effectively better than just using random trial states~\cite{yung2014introduction}. These advancements have served as inspiration for our research.

In this paper, we propose a classical-assisted quantum ground state preparation method for quantum many-body systems, by combining tensor network states~\cite{verstraete2008matrix,orus2014practical,czarnik2019time} and Monte Carlo ~\cite{wang2011monte} sampling as a heuristic method to prepare a trial state with a non-trivial overlap with the genuine ground state. 
Our method demonstrates a polynomial improvement in the scaling of overlap between the trial state and genuine ground state compared to using random trial states, reducing from $2^{0.5L}$ to $2^{0.5\delta L}$ and $0<\delta<1$. Notably, $\delta$ is only 0.37 in our numerical test, which implies that our time complexity exhibits a slower growth rate in relation to $L$.
On the classical aspect, we employ the simple update (SU) imaginary Time Evolution Block Decimation (iTEBD) algorithm~\cite{vidal2003efficient} which exhibits the efficient performance of gaining access to the ground state of high fidelity~\cite{orus2009simulation}.
Despite the high overlap of the entire TNS, preparing it as a trial state leads to high time complexity~\cite{schwarz2012preparing}.
Excitingly, a numerical investigation of spin-$1/2$ $J_1$-$J_2$ Heisenberg model suggests that the Hamiltonian ground state exhibits a phenomenon that ground state wavefunction has a localized distribution, shown in Appendix~\ref{appendix1}. Therefore, implement Monte Carlo sampling to generate sparse trial states with $M$ major components to reduce time complexity, and we provide the solution for obtaining the optimal value of $M$ that minimizes the time complexity.
And on the quantum aspect, we utilize the Faster Ground State Preparation (FGSP) algorithm~\cite{ge2019faster} which demonstrates the optimal time complexity for the known ground energy under suitable assumptions.

\begin{table}[t]
\caption{Symbol Table.}
\centering
\begin{tabular}{c l}
\toprule
Symbol & Meaning \\
\midrule
$H$ & Hamiltonian \\
$L$ & Size of the system, also means the number of qubits \\
$N$ & The dimension of Hamiltonian \\
$|\lambda_0\rangle$ & Genuine ground state of Hamiltonian \\
$\lambda_0$ & Lowest eigenvalue of Hamiltonian\\
$M$ & Number of major components for sampling \\
$T$ & Complexity of our proposed method\\
$m$ & Number of approximating the cosine of the Hamiltonian\\
$\epsilon$ & Given precision to the genuine ground state  \\
$\Delta$ & Minimum value for the spectral gap of Hamiltonian  \\
$\Phi$ & Gate cost for preparing a trial state  \\
$\Lambda$ & ``Base cost'' of the Hamiltonian simulation  \\
$f_1$ & Fidelity between PEPS and genuine ground state \\
$f_2$ & Fidelity between trial state and genuine ground state \\
$\alpha$ & Exponential factor of $f_2$ and $L$ \\
$\beta$ & Exponential factor of $f_1$ and $L$ \\
$\gamma$ & Exponential factor of $M$ and $L$ \\
\bottomrule
\end{tabular}
\end{table}

In the numerical test, we focus on the two-dimensional spin-$1/2$ $J_1$-$J_2$ Heisenberg model which is a popular strongly correlated quantum many-body model that describes the magnetic behavior of half-filled Hubbard model in large-$U$ limit.
The numerical test demonstrates a polynomial improvement in scaling of overlap compared to just using random trial states. Moreover, our method has the potential to be extended to other tensor network methods and quantum algorithms. For instance, they can be replaced by variational methods~\cite{Low_2010,reiner2019finding}, the Density Matrix Renormalization Group (DMRG)~\cite{white1992density,danshita2009bose,vidal2004efficient} or near-optimal quantum ground state preparation~\cite{lin2020near}.

The outline of the remainder of the paper is as follows: in Sec.~\ref{section2}, we provide the theoretical background of tensor network states. In Sec.~\ref{section3}, we elaborate on the technical details of the method and provide the solution for obtaining the optimal value of $M$ that minimizes the time complexity. In Sec.~\ref{section5}, a series of numerical tests demonstrate the feasibility of our algorithm and the effectiveness of polynomial acceleration. In Sec.~\ref{section6}, we discuss future research prospects and outlooks.
Collectively, our findings demonstrate that our approach presents the potential in applications of ground state preparation problems.

\section{Tensor Network States}
\label{section2}
Tensor network states  are a class of mathematical representations for entangled quantum states in many-body physics systems~\cite{verstraete2008matrix,orus2014practical,czarnik2019time}, including matrix product state (MPS) for one-dimensional lattices and pair-entangled projected state (PEPS) as its two-dimensional extension. In this section, we focus on PEPS and provide a concise overview of its theoretical foundations.

Consider a quantum state denoted as $|\Psi\rangle$, composed of $L=L_{1}\times L_{2}$ spins represented by the basis states $|s_{i,j}\rangle$. PEPS is defined on a two-dimensional lattice with $L$ sites, where each site possesses a physical degree of freedom denoted as $s$. The general quantum state of this system can be expressed in the following form:
\begin{equation}
\label{equ30}
|\Psi\rangle=\sum_{s}C_{s_{1,1},s_{1,2},...,s_{L_{1},L_{2}}}|s_{1,1},s_{1,2},...,s_{L_{1},L_{2}} \rangle
\end{equation}
where $C_{s_{1,1},s_{1,2},...,s_{L_{1},L_{2}}}$ is the amplitude of the wavefunction over the basis, which satisfies $\sum_{\{s\}}|C_{s_{1,1},s_{1,2},...,s_{L_{1},L_{2}}}|^2=1$.
The dimension of the Hilbert space is given by $N=d^{L}$, which exhibits an exponential increase with the size of the system. Here, $d$ represents the physical dimension at each site of the system. For the sake of simplicity, we assume a uniform dimension of $d=2$ for all sites.

However, as the system size $L$ increases, it becomes infeasible to represent all the amplitudes explicitly. TNS gives an approximated representation of the wavefunction by representing it as the contraction of individual tensors. For example, PEPS on a 2D lattice is constructed as below. Each site in the two-dimensional lattice, identified by its coordinates $(i,j)$, is assigned a tensor denoted as $A[i,j]^{S_{i,j}}_{d_{i,j,1}d_{i,j,2},...}$, where $d_{i,j,1}d_{i,j,2},...$ represents virtual indices corresponding to the virtual legs that connect with neighboring sites.
The PEPS can be expressed as:

\begin{equation}
|\Psi\rangle=\frac 1Z\sum_{\{s\}}C\Big[\prod_{i,j}A[i,j]^{S_{i,j}}_{d_{i,j,1}d_{i,j,2},...}\Big]|s_{1,1},s_{1,2},...,s_{m,n}\rangle
\end{equation}
where $C$ means to contract all connected virtual legs, $Z$ is the normalization constant and $A$ denotes the tensor. 

\begin{figure}[t]
 \centerline{
	\includegraphics[width=0.25\columnwidth]{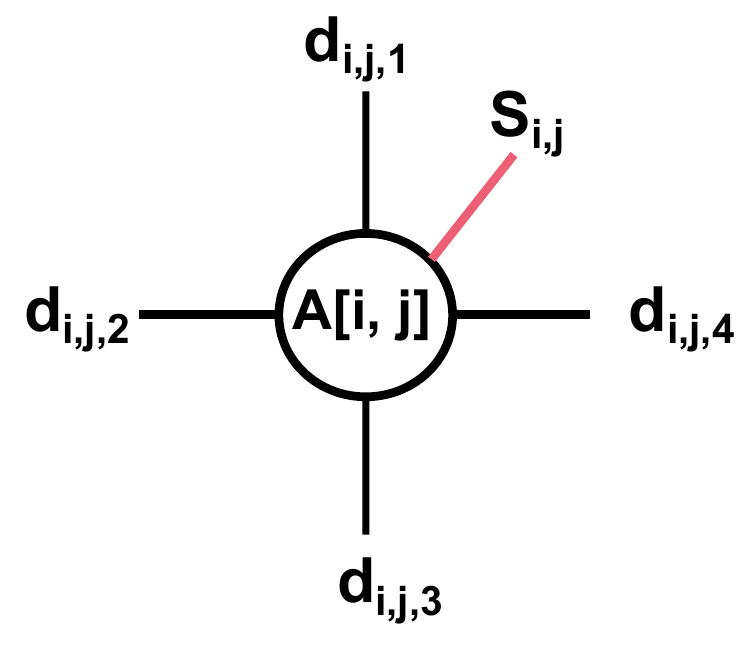}
 }
\caption{The graphical representation of tensor $A[i,j]^{S_{i,j}}_{d_{i,j,1}d_{i,j,2},...}$. Every leg denotes a tensor index. The red line represents the physical leg.}
\label{tns_1}
\end{figure}

\begin{figure}[t]
\centering
\subfigure[4 × 4 square lattice] { \label{fig:a1}
\includegraphics[width=0.39\columnwidth]{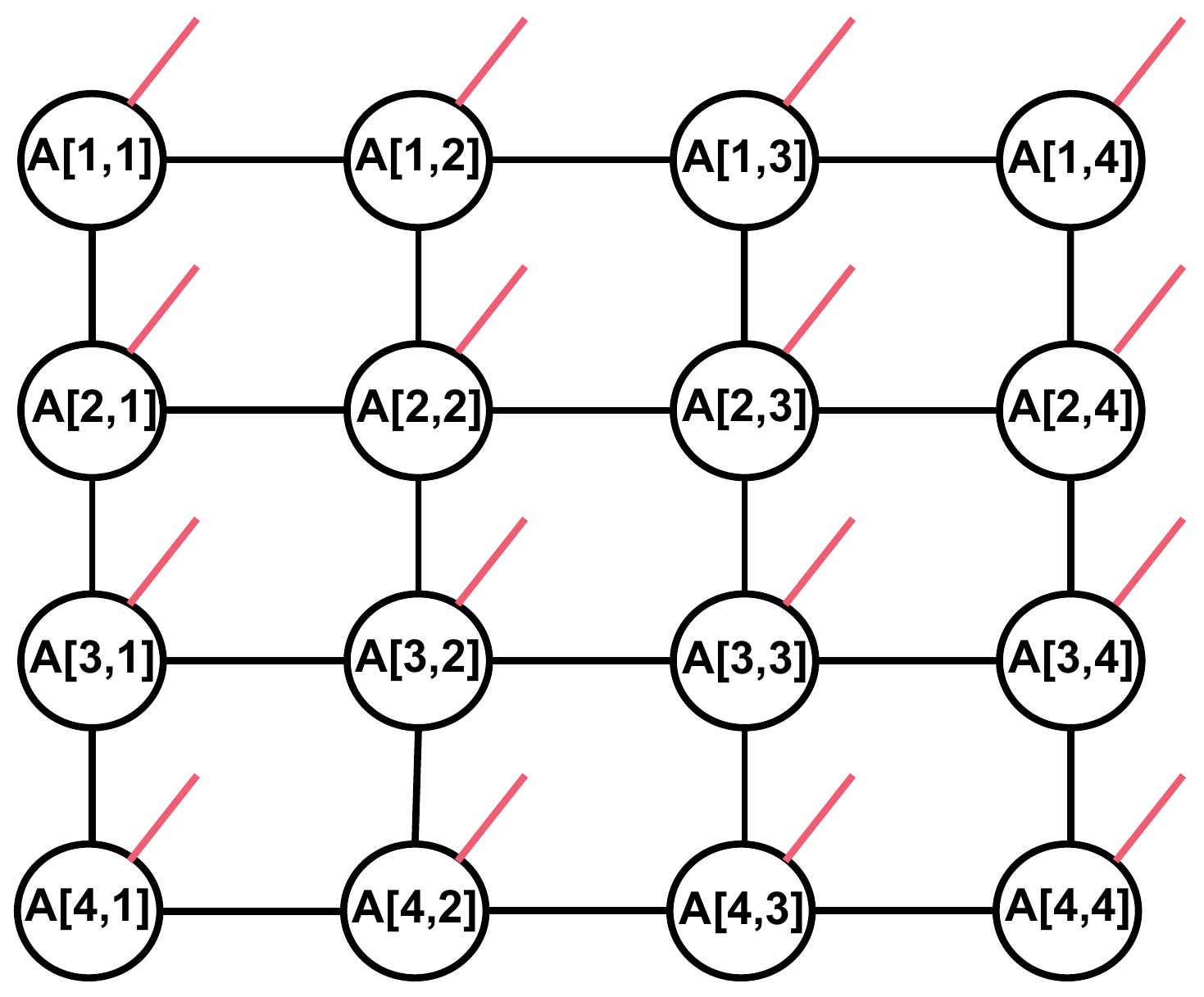}
}
\subfigure[4 × 4 triangular lattice] { \label{fig:b1}
\includegraphics[width=0.42\columnwidth]{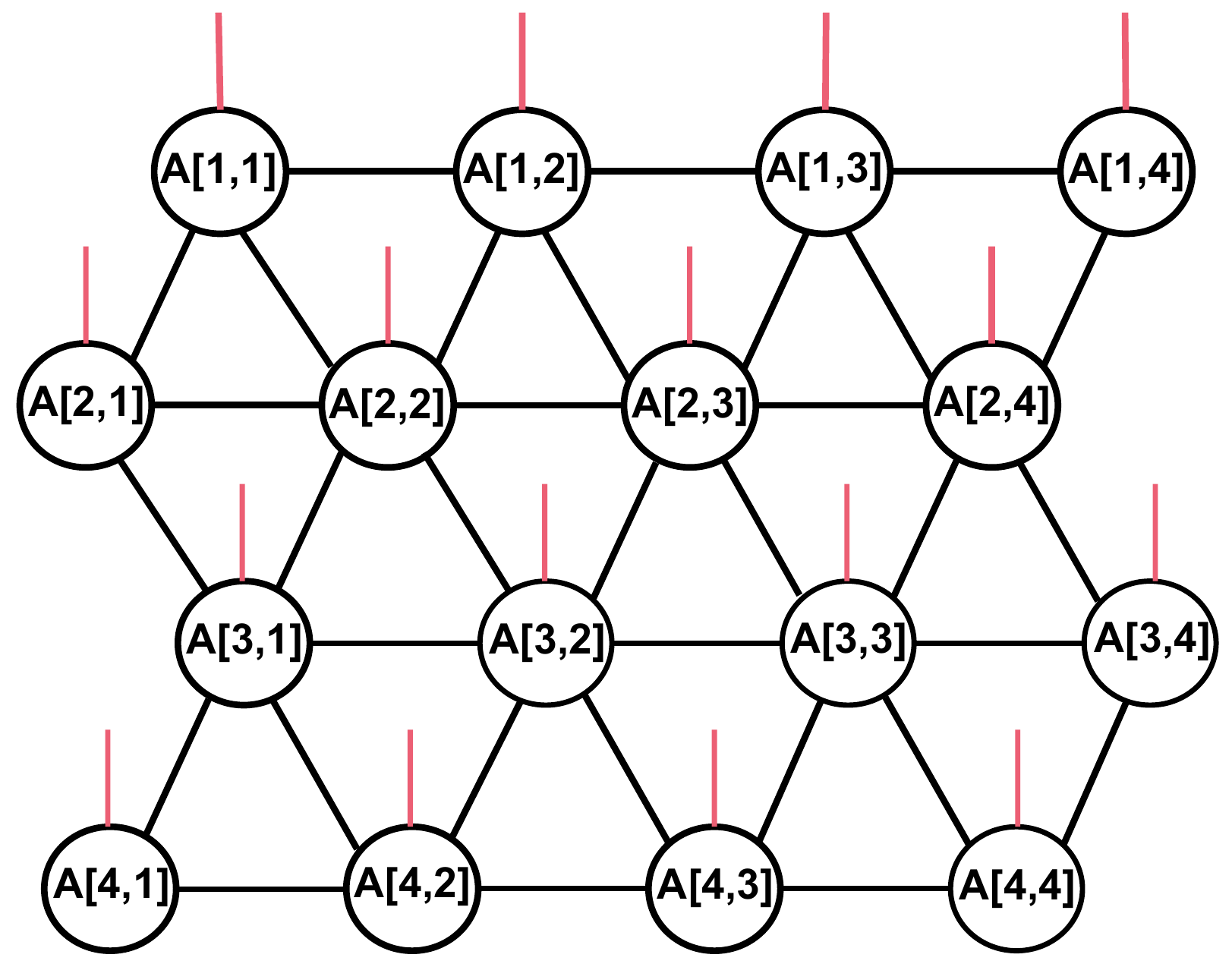}
}
\caption{(a) A PEPS on a $4\times4$ square lattice, where the tensors have a maximum of 4 virtual legs. (b) A PEPS on a $4\times4$ triangular lattice, where the tensors have a maximum of 6 virtual legs.
Each circle corresponds to a site in the lattice. The free legs are the physical legs.}
\label{tns}
\end{figure}

The exact form of PEPS is determined by the topology and the size of the lattice. For instance, consider the tensor $A[i,j]$ on a square lattice, which possesses 5 legs. This tensor can be visualized as a vector with dimensions $[D, D, D, D, d]$, where the first four legs represent the virtual indices, and the last leg corresponds to the physical index. The bond dimension $D$ denotes the dimension of virtual legs and $d$ denotes physical indices, establishing $D$ as the control parameter for the state. In order to represent the tensor more vividly, we use the graphical notation \cite{schollwock2011density,schollwock2011density2} to denote $A[i,j]$ and PEPS. Fig.~\ref{tns_1} graphically represents the single tensor, and Fig. \ref{tns} shows the PEPS on square and triangular lattices.

\section{Method}
\label{section3}

\subsection{Framework}
The core of our method is merging tensor network methods and Monte Carlo sampling to prepare a trial state with non-trivial overlap with the genuine ground state for certain quantum ground state preparation methods in quantum many-body systems. 
And the method presented here is based on the observation that the ground state of spin-$1/2$ $J_1$-$J_2$ Heisenberg model exhibits a phenomenon that ground state wavefunction has a localized distribution, shown in Appendix~\ref{appendix1}. As depicted in  Fig.~\ref{fig1}, our method contains three main steps: classical, trial state, and quantum, which is structured as follows:
\begin{itemize}
  \item [1.] 
  We obtain the rough ground state $|\varphi_{0}\rangle$ on classical computers through certain tensor networks methods like iTEBD, DMRG, and variational methods~\cite{Low_2010}.      
  \item [2.]
  Rather than encoding entire $|\varphi_{0}\rangle$ into the trial state in a quantum computer, we select the major components of the $|\varphi_{0}\rangle$ by Monte Carlo sampling~\cite{wang2011monte}. 
  \item [3.]
  Utilize the trial state in certain quantum ground state preparation algorithms.
\end{itemize}

\begin{figure}[!h]
 \centerline{
	\includegraphics[width=0.7\columnwidth]{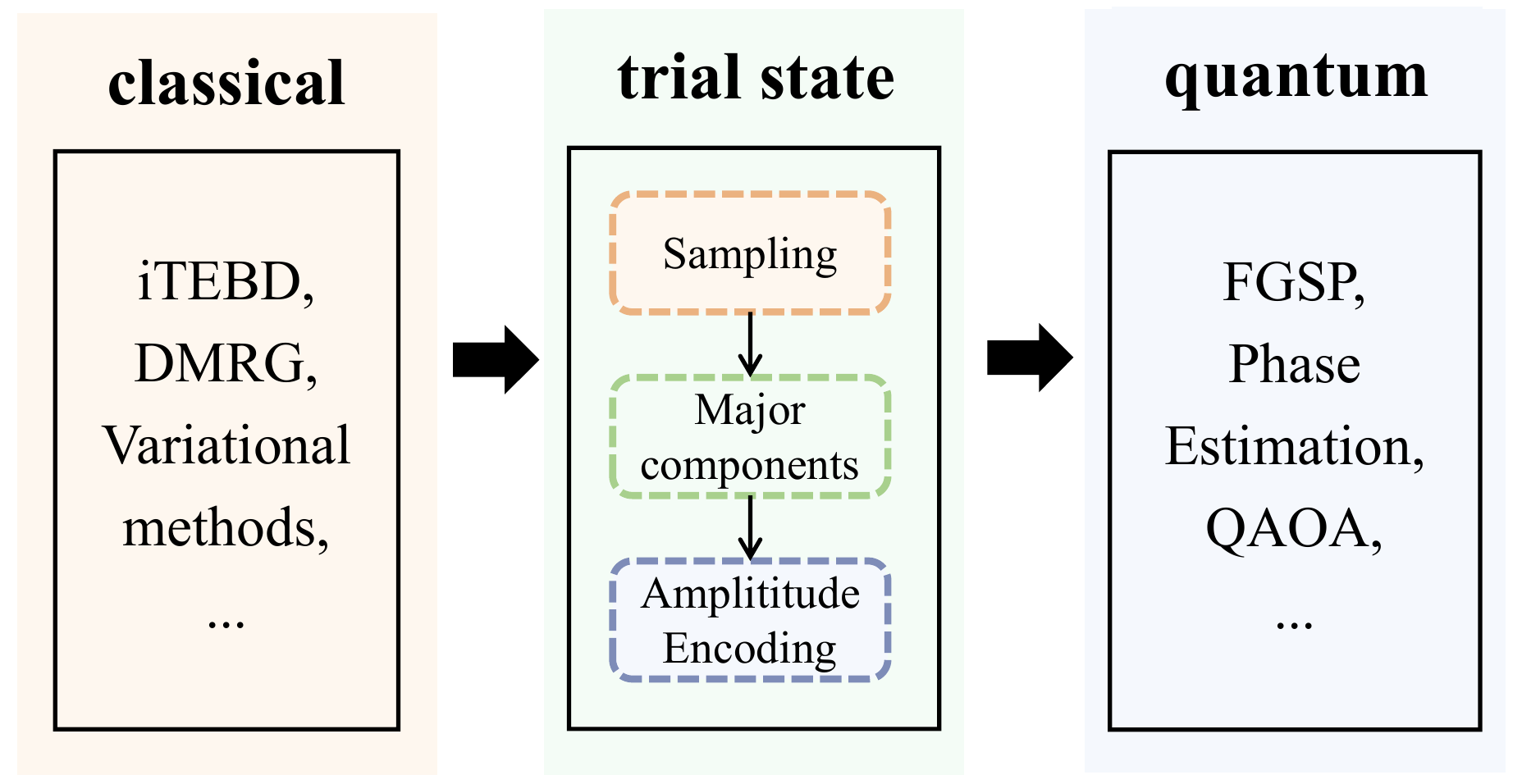}
 }
\caption{An illustration of the classical and quantum steps of our method: (1) Obtain the approximate state from the classical ground state preparation method, like iTEBD, DMRG, and variational methods; (2) Prepare the trial state by sampling major components; (3) Through amplitude encoding, the trial state is utilized in quantum ground state preparation algorithms, including FGSP, phase estimation algorithms, and QAOA method. Finally, we can obtain the ground state by measurement.}
\label{fig1}
\end{figure}

The candidates for our classical-assisted method are the iTEBD method with ansatz TNS and FGSP respectively. 
The iTEBD method exhibits the efficient performance of gaining access to the ground state of high fidelity and lower time complexity than other tensor network methods. The FGSP method scales exponentially better in the allowed error to the genuine ground state, and polynomially better with the spectral gap and the overlap compared to using phase estimation. 

To describe this method more concretely, it is convenient to assume that $H$ is an $N\times N$ Hermitian matrix and $N = 2^{L}$. 
We are interested in gaining access to the genuine ground state $|\lambda_{0}\rangle$ and ground energy $\lambda_{0}$ of $H$. 
In conjunction with the description in~\cite{ge2019faster}, we present the specific implementation steps outlined in Algorithm~\ref{A1}. Here, we modify some symbols and prepare three requirements in advance: Hamiltonian $H$, the number of major components $M$, and the value of parameter $m$. First, we acquire rough normalized tensor network state $|\varphi_{0}\rangle$ and energy $E_{0}$\footnote{In quantum many-body systems, the average lattice energy is computed as $E=\langle \varphi |H| \varphi \rangle /L$, where $\varphi$ represents the tensor and $L$ corresponds to the size of system.} by the iTEBD method on the classical computer. 
Second, we employ the Monte Carlo method to sample the major components of $|\varphi_{0}\rangle$. 
The number of sampling sweeps $T$ is related to the size of the system. 
The sparse trial state can be represented as 
$|\varphi_{1}\rangle=\sum_{j=0}^{M-1}{c_{a_j}|a_j\rangle}$, where $|a_j\rangle$ denotes major component and $c_{a_j}$ is the corresponding amplitude. 
Third, we utilize the Faster Ground State Preparation in 3-5 steps when the energy is known. Here, we denote the overlap of trial state and genuine ground state as $|\phi_0|=|\langle\lambda_{0}|\varphi_{1}\rangle|$.

\begin{algorithm}
 \caption{Algorithm for ground state preparation combines simple update iTEBD with TNS and FGSP.}
 \label{A1}
 \begin{algorithmic}[1]
        \REQUIRE Hamiltonian $H$; The number of major components $M$; The number of approximating the cosine of the Hamiltonian $m$.
        \STATE Obtain rough normalized ground state $|\varphi_{0}\rangle$ and energy $E_{0}$ by iTEBD method;
        \STATE Prepare the trial state $|\varphi_{1}\rangle=\sum_{j=0}^{M-1}{c_{a_j}|a_j\rangle}$ by sampling $M$ major components from TNS;
        \STATE Approximate ${\cos}^{m}H$ as a linear combination of terms of the form $e^{-iHt_k}$;
        \STATE Using the techniques in Ref.~\cite{childs2017quantum}, we implement this linear combination on $|\varphi_{1}\rangle$ using Hamiltonian simulation and the Linear Combination of Unitaries (LCU).
        \STATE Using amplitude amplification to obtain the ground state.
\end{algorithmic}
\end{algorithm}
The detail of our hybrid methods and the time complexity of which will be described in the following sections. Since the quantum component of our method serves as an application of FGSP, we have provided a detailed introduction to the FGSP method in Appendix \ref{appendixFGSP}. Therefore, the following sections are classical, trial state, and complexity analysis.

\subsection{Classical Ground State Preparation}

We utilize the iTEBD method with the ansatz TNS as an example for quantum many-body systems. Tensor network state has been demonstrated to be a remarkably effective tool for studying the physics of strongly correlated many-body systems in one and two dimensions. The iTEBD method is computationally efficient, requiring a lower computational cost.

The detailed structure of iTEBD method involves three steps.
First, initialize the TNS representation of wavefunction, which involves setting up the tensor network structure and assigning random values to the tensors.    
Second, run several steps of time evolution which is described in~\cite{dong2018tnspackage}. 
Third, calculate the energy as an evaluation criterion for the ground state.

Our code is built on the TNS Package~\cite{dong2018tnspackage}. We have tested our methods on the two-dimensional spin-$1/2$ $J_1$-$J_2$ Heisenberg model with bond dimensions $D=6$ and ran 300 times of each step length which decreases gradually from 0.1 to 0.00001. 
Moreover, selecting appropriate values of $D$ and steps can enable the extension of the algorithm to higher-dimensional scenarios.

\subsection{Prepare Trial State} 
How to prepare the trial state from TNS on classical computer$?$
It is worth noting that despite the high overlap of the entire tensor network state, utilizing it as a trial state results in high time complexity. 
In light of this, our algorithm takes inspiration from the observation that the ground state of the spin-1/2 $J_1$-$J_2$ Heisenberg model exhibits a phenomenon that ground state wavefunction has a localized distribution.
To capture the major components efficiently, we employ the Monte Carlo sampling technique, aligning with this motivation.

\begin{figure}[!h]
 \centerline{
	\includegraphics[width=0.9\columnwidth]{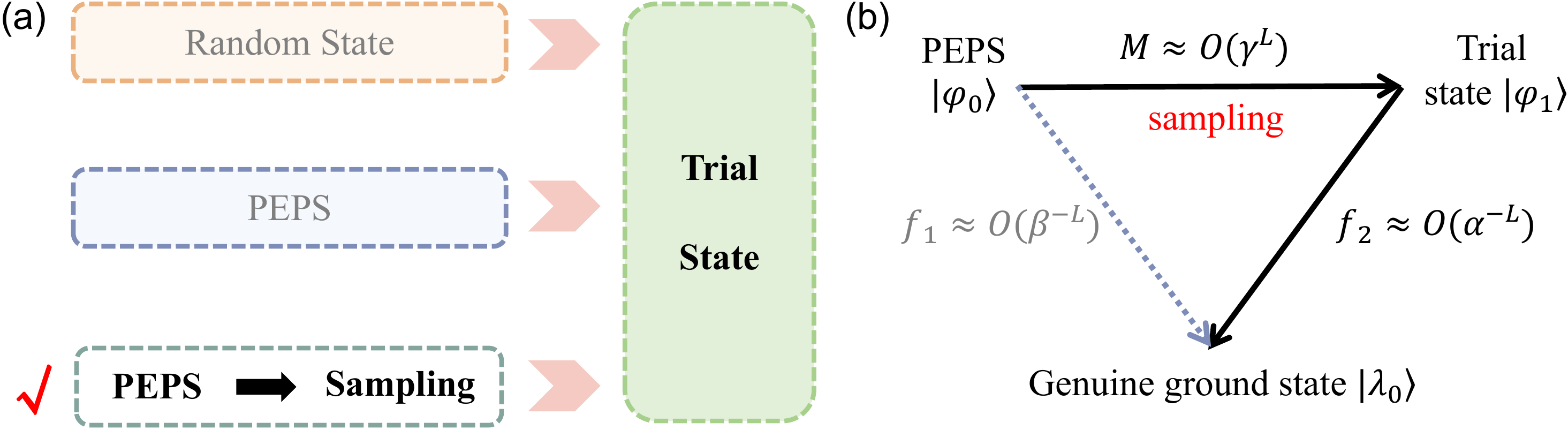}
 }
\caption{Three trial state preparation approaches and illustration of our preparation route. In Fig.(a), combing PEPS and MC sampling surpasses the other algorithm using random state or entire PEPS in terms of efficiency under the same conditions in Sec.~\ref{section5}. In Fig.(b), the three vertices of the triangle represent the PEPS, trial state, and genuine ground state. The blue dotted arrow shows the exponential decay of fidelity $f_1$ between PEPS and genuine ground state as $L$ increases, with $\beta$ ($1 < \beta < 2$) as the exponential factor.  When $\beta$ equals 1, the fidelity remains constant regardless of $L$, which is ideal. Conversely, $\beta$ equals 2 indicates using a random trial state. Similarly, the black arrow signifies the fidelity $f_2$ between our trial state and the genuine ground state, characterized by $\alpha$ ($1 < \alpha < 2$).
Additionally, assuming exponential growth of the number of samples $M$ with $L$, characterized by the exponential factor $\gamma$, we can derive $\alpha\approx\beta/\gamma$ from Eq.~(\ref{fidelity1}).}
\label{trial_state}
\end{figure}

Three distinct approaches of trial state preparation are illustrated in Fig.~\ref{trial_state}(a). Among these, our proposed algorithm represents the third approach, while the remaining two serve as comparison methods in this paper.
As depicted in Fig.~\ref{trial_state}(b), we focus on the fidelity relationship among the PEPS, trial state, and genuine ground state, where fidelity represents the square of the overlap.

The blue dotted arrow indicates the preparation route of random states and PEPS in Fig.~\ref{trial_state}(a). We assume this fidelity $f_1$ between PEPS and genuine ground state decays exponentially as $L$ increases, characterized by an exponential factor $\beta$ where $1 < \beta < 2$.
In the case where $\beta$ is equal to 1, the fidelity remains constant regardless of the value of $L$ which is ideal for us. Conversely, when $\beta$ equals 2, it indicates that the trial state is a random state. 
Inspired by the observation that the ground state distribution of the spin-1/2 $J_1$-$J_2$ Heisenberg model seeming exhibits Normal distribution, we assume a Gaussian error relationship function ($\operatorname{erf}$) between the number of major components $M$, size $L$ and $f_1$:
\begin{equation}
\label{fidelity1}
f_1(M,L)=g\cdot{\operatorname{erf}}(a\frac{M}{\beta^L}+e)+h=\frac{2g}{\sqrt{\pi}} \int_{0}^{a\frac{M}{\beta^L}+e}e^{-t^2}dt + h,
\end{equation}
where $a,e,g,h$ are parameters added for modification. We have performed a fitting of the spin-1/2 $J_1$-$J_2$ Heisenberg model in the case of $L=4\times2$ in Appendix~\ref{appendix1}, and the result demonstrates a well fit. As $\beta_L$ varies across different sizes of $L$, we assume an upper bound $\beta = \max\{\beta_L\}$.

The black arrows symbolize our preparation route, which highlights our two approximations of the genuine ground state. The first approximation involves the optimization of projected entangled-pair states, while the second approximation utilizes the MC sampling method. After MC sampling, the fidelity $f_2$ between our trial state and genuine ground state decays exponentially as $L$ increases, characterized by an exponential factor $\alpha$ where $1 < \alpha < 2$. Additionally, given that $M$ is known after sampling, we assume the relationship about $f_2$ and $L$ as 
\begin{equation}
\label{fidelityf2}
f_2(L)=\frac{a}{\alpha^{L}}+e,
\end{equation}
where $a,e$ are modification parameters. In this paper, we treat these irrelevant modifier parameters as undifferentiated, despite their distinct values. Next, we illustrate the reasonableness of the assumption of this model.
It is noteworthy that observation demonstrates that $M$ seemingly can be a function of the exponential growth of $L$, characterized by an exponential factor $\gamma$. 
\begin{equation}
\label{fitting_gamma}
M(L) = a\gamma^L+e,\\
\end{equation}
where $a,e$ are modification parameters. By substituting this equation into Eq. (\ref{fidelity1}), we can treat $L$ as an independent variable and obtain 
\begin{equation}
\label{fidelity2}
f_2(L) = \widetilde{\mathcal{O}}((\gamma/ \beta)^L),
\end{equation}
which demonstrates our assumption is correct and $\alpha\approx \beta/ \gamma$.

Building upon these three models, we observed the key of our method is gaining the model of $M$ and $L$. The number of major components $M$ ranges theoretically between 0 and $2^L$ for a system of scale $L$. However, a smaller $M$ results in lower fidelity, while a larger $M$ leads to higher time complexity for amplitude encoding. We will explore deeper into the analysis of the optimal complexity by exploring the values of $M$ and $L$, along with the corresponding conditions.

\subsubsection{Monte Carlo Sampling}
The Monte Carlo sampling method is a tool that uses a probabilistic approach to sample the  basis vectors (which we call configurations) according to their possibility with respect to a given wavefunction.  Here, we represent the wavefunction by $|\varphi_0\rangle=\frac{1}{Z}W({\bm{S}})|\bm{S}\rangle$, where $\bm{S} =s_{1,1} s_{1,2} ...$ denotes a configuration and $Z=\sum_{\bm{S}}|W^2(\bm{S})|$ denotes normalization coefficient.
The probability for each configuration $|\bm{S}\rangle$ to be sampled is $\frac 1Z|W^2(\bm{S})|$. Therefore the more important configurations are more likely to be sampled within the limited number of samples. The detailed process of MC sampling on PEPS is as follows:
\begin{itemize}
  \item [1.] 
  Initial a randomly generated PEPS configuration denoted as $\bm{S_0}$. 
  \item [2.]
  Randomly select a lattice position and flip the corresponding spin to generate a new configuration as $\bm{S_1}$. Calculate the squared amplitudes before and after the update, $W^2 (\bm{S_0})$ and $W^2(\bm{S_1})$ respectively.
  \item [3.]
  Generate a uniformly distributed random value $r$ ranging from 0 to 1. If $p_0 = \frac{W^2(\bm{S_1})}{W^2(\bm{S_0})} > r$, we accept the update and consider the new configuration as the starting point for the next step. Conversely, this update is rejected.
  \item [4.]
  By iteratively performing the update and acceptance steps, we construct a Markov chain where each step generates a configuration of the PEPS.
  
\end{itemize}

In our numerical test, we perform Monte Carlo sampling on $|\varphi_{0}\rangle$. The number of MC samples ranges from $5,000$ to $50,000$ as the system size varies from $L = 8$ to $L = 20$.

\subsection{Complexity analysis}
\label{section4}
Our proposed algorithmic analysis includes two components of time complexity: classical and quantum. The classical analysis characterizes the time complexity of the imaginary time evolution and sampling performed on a classical computer. The quantum analysis $T_{q}$ consists of a time complexity analysis of quantum circuits involving trial state preparation and FGSP.

\subsubsection{Classical Algorithm Analysis}

The time complexity analysis of the classical part includes both the imaginary time evolution and the computational energy, which is mainly related to the virtual bond dimension $D$, samples $M$, and size $L$ of the tensor network states. 

For optimization, we use an imaginary time evolution with a simple update algorithm~\cite{jiang2008accurate}, in which the environment of tensors is approximated by the product of some diagonal matrices. In this method, the tensors are updated site by site by singular value decomposition, and the bond dimensions are truncated back to $D$. By combining QR/LQ decomposition, when dealing with the nearest neighbor (NN) interactions~\cite{wang2011monte}, the computational cost in the update process is greatly reduced to $\mathcal{O}(D^{5})$. 

For calculating energy and $W(S)$ once, the computational scaling cost is as high as $\mathcal{O}(D^{10})$~\cite{verstraete2004renormalization,verstraete2008matrix} when strictly contracting the whole tensor network in square lattices with open boundary conditions (OBC). Using the Monte Carlo sampling technique, the computational cost is $\mathcal{O}(D^{6})$~\cite{liu2017gradient}.

Our experience shows that in large systems such as $8\times8$, the configurations of Markov chains are basically unduplicated and of the same order of magnitude as $M$. Therefore, considering $L$ site shrinkages and $M$ executions at a time, the total computational cost is $\mathcal{O}(MD^{6}L)$. 

\subsubsection{Quantum Algorithm Analysis}
According to Ref.~\cite{ge2019faster}, $\Lambda$ is the ``base cost'' of the simulation, $\Delta$ is the known lower bound on the spectral gap of $H$ and $\Phi$ is preparation of a trial state. The total quantum gate complexity $T_{q}$ is
\begin{equation}
\label{equ4}     
T_{q}=\widetilde{\mathcal{O}}(\frac{\Lambda}{|\phi_0|\Delta}+\frac{\Phi}{|\phi_0|}),
\end{equation}
where $\Delta$ and $\Phi$ are necessary to analyze according to specific Hamiltonian, as different Hamiltonian properties can be simulated by choosing an appropriate method. 
In this paper, the cost of our state preparation $\Phi$ is $O(ML)$ which is related to the number of selected major components $M$ and the scale of the system $L$ according to~\cite{malvetti2021quantum,de2022double,gleinig2021efficient}. We will discuss $\Phi$ of specific Hamiltonian in Sec.~\ref{Hamiltonians_analysis}.

Ultimately, owing to the disparate computational efficiencies and temporal characteristics of classical and quantum computers, it is conjectured that the processing time of a single step on a classical computer is considerably shorter compared to that of a quantum computer. The comprehensive time complexity is subsequently derived by amalgamating the computational aspects of both classical and quantum components, thereby yielding the following expression:
\begin{equation}
\label{equ5} 
T=\widetilde{\mathcal{O}}(MD^6L+\frac{ML}{|\phi_0|}+\frac{\Lambda}{\Delta|\phi_0|}).
\end{equation}
It is clear from the equation that the total time complexity depends on the values of $M$ and $L$ since $|\phi_0|$ is determined by them.

\subsubsection{Optimality of the complexity}
\label{optimal_complexity}
In this section, our objective is to determine the optimal complexity of our method by choosing appropriate values for $M$ and $L$. We begin by delving into the process of obtaining the optimal $M$ in the case of known $L$.
Next, we provide the time complexity with three distinct trial state preparations and derive the conditions based on a comparison with the other two complexities.

First, we discuss the value of $M$ for optimal time complexity. As previously mentioned, we aim to treat $L$ as an independent variable, but it is necessary to get the model of optimal $M$ and $L$ by Eq.(\ref{fitting_gamma}). The key to how to balance the complexity of classical and quantum and achieving the optimal time complexity lies in the selection of an appropriate value of $M$, when in the case of fixed $L$. We expect to obtain the optimal $M$ by deriving $T$ with respect to $M$
\begin{equation}
\frac{\partial T}{\partial M}
= D^6L+{Lf_1(M)^{-1/2}}-(ML+\frac{\Lambda}{\Delta}){\frac{1}{2f_2(M)^{3/2}}\frac{\partial f_1(M)}{\partial M}}=0.\\
\end{equation}
As $M$ tends to 1, the derivative tends towards negative infinity.
As $M$ tends to be $2^L$, the derivative tends to be $D^6L$. Because as $M$ increases, $\partial f_1(M)/ \partial M$ and $1/f_1(M)$ decrease and $0 < 1/f_1(M) < 1$, the derivative is a curve that continues to grow from negative infinity. Therefore, the time complexity seeming can be seen as a continuous concave curve.
Thus, we get the value of $M$ when ignoring some items
\begin{equation}
M=(f_1(M)^{\frac{3}{2}}D^6+f_1(M))(\frac{\partial f_1(M)}{\partial M})^{-1}-\frac{\Lambda}{L}.
\label{equ12}
\end{equation} 
The optimal $M$ is obtained by drawing or solving analytically through Eq.~(\ref{equ12}) to optimize time complexity. Thus, we can obtain the exponential factor $\gamma$ by fitting the optimal value of $M$ with the corresponding value of $L$ using Eq.~(\ref{fitting_gamma}). Finally, we can think of $T$ as an exponential relation with respect to L:
\begin{equation}
\label{Tfinal}
T=\widetilde{\mathcal{O}}(\gamma^LD^6L+\frac{\gamma^L L+\Lambda}{\Delta \sqrt{\operatorname{erf}((\gamma / \beta)^L)}})\approx \widetilde{\mathcal{O}}((\gamma \beta)^{(L/2)}).
\end{equation}
The derivation process for this step is presented in Appendix~\ref{derivation}.

Second, we discuss the conditions when our proposed method is lower than the other two methods. Tab.~\ref{table1} displays the gate complexity associated with the three distinct methods employed for trial state preparation. The primary discrepancy arises from the complexity variation of the $\Phi$ cost and overlap, since the $\Lambda$ and $\Delta$ are same in one Hamiltonian. 
The random state can be generated by employing a random circuit of constant depth, denoted as $c$. For the sake of convenience, we assume the cost of state preparation is $\mathcal{O}(1)$ in the case of utilizing random state.
The PEPS preparation cost on a quantum computer in~\cite{schwarz2012preparing} is approximately $\widetilde{\mathcal{O}}(\frac{|V|^2|E|^2\kappa^2}{\epsilon\Delta}+|V|kd^6)$. Here, $G = (V, E)$ represents an interaction graph with a bounded degree and a predefined total order on $V$. The set ${A(v)}_{v\in V[t_2]}$ comprises injective PEPS projectors of dimension $d\times D^k$ associated with each vertex $v$ in $V$ up to the vertex $t_2$ (according to the total vertex order), forming a sequence of PEPS. The term $\kappa = \max_{v\in V}\kappa(A(v))$ denotes the largest condition number among all PEPS projectors.

\begin{table*}[!h]
\centering
\setlength{\belowcaptionskip}{0.2cm}
\caption{Gate complexity for three trial state preparations using FGSP method in the case when the ground energy is known beforehand to the required precision.}
\renewcommand\arraystretch{1.5}
\label{table1}
\begin{threeparttable} 
\begin{tabular}{l c c c}
\toprule
Preparation & Gates  & $\Phi$  & $|\phi_0|$
\\
\midrule
This paper    & $\widetilde{\mathcal{O}}(MD^6L+\frac{\Phi}{|\phi_0|}+\frac{\Lambda}{\Delta|\phi_0|})$  & $\mathcal{O}(ML)$  & $\mathcal{O}(\alpha^{-L/2})$\\
Random state  & $\widetilde{\mathcal{O}}(\frac{\Phi}{|\phi_0|}+\frac{\Lambda}{\Delta|\phi_0|})$ &$\mathcal{O}(1)$ & $\mathcal{O}(2^{-L/2})$\\
PEPS  & $\widetilde{\mathcal{O}}(D^6L+\frac{\Phi}{|\phi_0|}+\frac{\Lambda}{\Delta|\phi_0|})$ &$\widetilde{\mathcal{O}}(\frac{|V|^2|E|^2\kappa^2}{\epsilon\Delta}+|V|kd^6)$ & $\mathcal{O}(\beta^{-L/2})$\\
\bottomrule
\end{tabular}
 \begin{tablenotes}
    \footnotesize 
    \item[1] Exponential factor $\alpha$ and $\beta$ are equivalent to those in Fig.~\ref{trial_state}.
\end{tablenotes}
\end{threeparttable}
\end{table*}

Given that the state preparation complexity of PEPS incorporates the precision $\epsilon$, typically set to $\epsilon=10^{-4}$, we assume that the complexity of our proposed method is lower than PEPS state preparation. Therefore, we only discuss the conditions of $L$ when the time complexity of our proposed method is lower than the random state. We assume that the complexity of using random states is $O(2^{L/2})$, and compare it with our method in Eq.~(\ref{Tfinal}):
\begin{equation}
\frac{T^{'}}{T}=2^{(1-\log_2(\gamma\beta))\frac{L}{2}} > 1.\\
\label{equ11}
\end{equation}
The efficiency of our algorithm is higher than the random state when under the condition
\begin{equation}
\label{equ15}
\gamma\beta<2.
\end{equation}
Finally, by utilizing Eq.~(\ref{equ12}) and (\ref{equ15}), we can determine the optimal time complexity, where a larger value of $L$ corresponds to a more significant advantage on our method.

\section{Experiments}
\label{section5}
Our numerical test aims to demonstrate that our proposed method can obtain lower time complexity compared to the random trial state and entire PEPS preparation, by providing a trial state with iTEBD and MC sampling methods. 

We apply our algorithm to the spin-$1/2$ $J_1$-$J_2$ Heisenberg model on a square lattice, which is a popular strongly correlated quantum many-body model that describes the magnetic behavior of half-filled Hubbard model in large-$U$ limit. This model is important as it describes the magnetic background of cupric oxide which becomes a high-Tc superconductor under doping~\cite{lee2006doping,singh1999dimer,einarsson1995direct,schulz1996magnetic,ivanov1992frustrated}. On the other hand, this model is also a possible model that holds spin-liquid phase, which is arguably topologically ordered~\cite{wang2016tensor,hu2013direct,gong2014plaquette,jiang2012spin,yu2012spin,liu2022gapless}.

In this section, we begin by introducing the theoretical knowledge and significance of the test model. Then, we perform a Hamiltonian simulation time complexity analysis of the test model. Finally, we give an analysis of our experimental results.

\subsection{Model}
The Hamiltonian of the spin-$1/2$ $J_1$-$J_2$ Heisenberg model on a square lattice reads
\begin{equation}
\label{equ7}
H=J_{1}\sum_{\langle i,j\rangle}\bm{S_{i}}\cdot\bm{S_{j}}+J_{2}\sum_{\langle\langle i,j\rangle\rangle}\bm{S_{i}}\cdot\bm{S_{j}}, \quad (J_{1}, J_{2} > 0),
\end{equation}
where $i,j$ means two-dimensional coordinates, $\langle ... \rangle$ denotes nearest neighbors,  $\langle\langle ...\rangle\rangle$ denotes the next-nearest neighbor, and $\bm{S_{i}}=\frac{1}{2}\mathop{\sigma_i}\limits ^{\rightarrow}$ represents the spin operator at site $i$. $J_1$ and $J_2$ represent the strength of nearest neighbor and next-nearest neighbor interaction respectively.

The spin-$1/2$ $J_1$-$J_2$ Heisenberg model has a rich phase diagram with varied $J_2/J_1$ value~\cite{liu2022gapless}. When $J_2/J_1< 0.45$ the model is in a magnetic-ordered Neel phase. In the region,  $0.45 < J_2/J_1 < 0.56$, the model is in a possibly gapless spin liquid phase. In the region,  $0.56 < J_2/J_1 < 0.61$, the model is in a valence bond solid phase. In the region,  $J_2/J_1 > 0.61$, the model is in a magnetic-ordered stripe phase~\cite{liu2022gapless}.

\subsection{Hamiltonian Simulation Analysis}
\label{Hamiltonians_analysis}
In this paper, we apply the algorithm presented in Ref.~\cite{low2019hamiltonian} which does not require the Hamiltonian to be expressed in any particular form. It is designed to work efficiently for both sparse and non-sparse Hamiltonians, and it can simulate the evolution of Hamiltonians with arbitrary time-dependent coefficients.
We employ Corollary 16 in~\cite{low2019hamiltonian} which is called Hamiltonian Simulation of a Linear Combination of Unitaries. This method proposes that given a Hamiltonian denoted as $H=\sum_{j=1}^d \alpha_j U_j$ where $\Vert \vec{\alpha}\Vert_1=\sum_{j=1}^d|\alpha_j|$ and $d$ represents the number of components, the ``base cost'' complexity is $\mathcal{O}(\Vert \vec{\alpha}\Vert_1)$. 

For our test model Hamiltonian, we can derive $\alpha_j=\{J_1, J_2\}$ and $U_j=\frac{1}{4}(\sigma^x_{j_1}\sigma^x_{j_2}+\sigma^y_{j_1}\sigma^y_{j_2}+\sigma^z_{j_1}\sigma^z_{j_2})$ where $j$ refers to the set of two-dimensional coordinates of the nearest and next-nearest neighbors and the set has two components $j_1$ and $j_2$. For example, $j=0$ denotes the set of $\{j_1=(0,0),j_2=(0,1)\}$.
Considering the structure of lattices in Fig.~\ref{tns}, we can get $d=\mathcal{O}(L)$ because the number of sets of nearest and next-nearest neighbors is proportional to $L$.

Finally, our proposed method complexity of spin-$1/2$ $J_1$-$J_2$ Heisenberg model is
\begin{equation}
T=\widetilde{\mathcal{O}}(MD^6L+\frac{ML}{|\phi_0|}+\frac{L}{{\Delta}|\phi_0|}).
\end{equation}

\subsection{Results}

In order to obtain fidelity $f_2$ of trial state and genuine ground state, it is crucial to acquire the genuine ground state via the Lanzcos algorithm in advance. The experimental bond dimension is $D=6$. In the part of imaginary time evolution on a classical computer, we do not necessarily need to achieve the best accuracy and could choose smaller steps. We ran 300 times each step length which decreases gradually from 0.1 to 0.00001.

We determine the optimal value of $M$ through the construction of a graph illustrating the time complexity. This procedure consists of two steps. First, we plot a curve representing the time complexity as a function of $M$, aiming to identify the minimum value across different $L$ values. The outcomes of the model testing are recorded in Tab.~\ref{table2}, which presents the approximate energy ($E_0$) of TNS alongside the corresponding optimal $M$ values for different system sizes ($L$). 

\begin{table*}[h]
\centering\setlength{\belowcaptionskip}{0.2cm}
\caption{Results of various sizes of testing models, which show the approximate energy ($E_0$) of TNS and optimal point ($M$) which leads to the lowest time complexity for various system sizes ($L$).}
\renewcommand\arraystretch{1.2}
\tabcolsep=0.4cm
\label{table2}
\begin{threeparttable} 
\begin{tabular}{c c c c c c c c}
\toprule
\multicolumn{2}{c}{\multirow{2}{*}{$L$(2D)}} & \multicolumn{2}{c}{$J_2/J_1=0$} & \multicolumn{2}{c}{$J_2/J_1=0.5$} & \multicolumn{2}{c}{$J_2/J_1=0.7$}\\
\cline{3-8}
\multicolumn{2}{c}{} & $E_0$ & $M$ &$E_0$ & $M$ & $E_0$ & $M$ \\
\midrule
\multicolumn{2}{c}{\textbf{4$\times $2}}    & -0.53487 & 4     & -0.45374  & 4    & -0.41321 & 7 \\ 
\multicolumn{2}{c}{\textbf{4$\times $3}}   & -0.55097 & 9    & -0.45571  & 11    & -0.40416 & 15 \\
\multicolumn{2}{c}{\textbf{4$\times $4}}    & -0.56680 & 16    & -0.46252  & 18    & -0.41655 & 28 \\
\multicolumn{2}{c}{\textbf{4$\times $5}}    & -0.57519 & 23    & -0.46412  & 32    & -0.41635 & 50 \\
\bottomrule
\end{tabular}
 \begin{tablenotes}
    \footnotesize 
    \item[1] In the 2D case, we set the size of the square lattice to be fixed with the first digit as 4.
\end{tablenotes}
\end{threeparttable}
\end{table*}

\begin{figure}[!h]
 \centerline{
	\includegraphics[width=0.55\columnwidth]{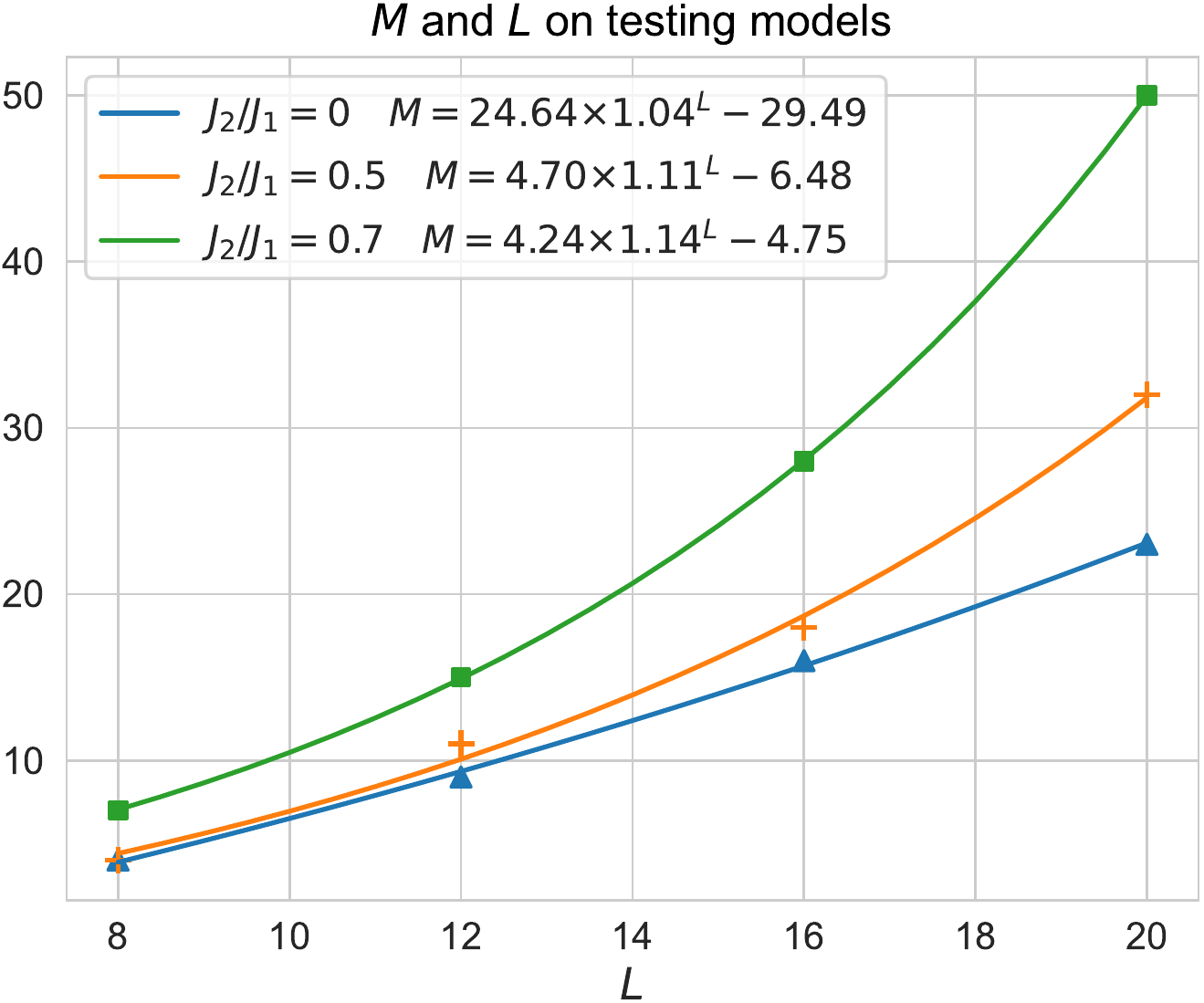}
 }
\caption{
The relationship between optimal $M$ and $L$ of the spin-$1/2$ $J_1$-$J_2$ Heisenberg model where $J_2/J_1=0$, 0.5 and 0.7. The fitted equation is $M = a\gamma^L+e$, where factor $\gamma$ can be as low as 1.04 in the case of $J_2/J_1=0$.}
\label{fig4}
\end{figure}

Second, fit $M$ versus $L$ according to the optimal value. The experimental figures for this part are placed in the appendix.  We generate curves representing the optimal values of $M$ and $L$ through the fitting Eq.~(\ref{fitting_gamma}), as illustrated in Fig.~\ref{fig4}. It is noteworthy that the base $\gamma$, characterizing the scaling behavior, exhibits values as low as 1.04. Additionally, the fitting yields a maximum value of 1.24 for $\beta$. By considering Eq.~(\ref{Tfinal}), we can get the exponential factor of $\gamma\beta\approx1.29$ from our numerical test. In other words, we achieve a reduction in time complexity from $2^{0.5L}$ to $1.29^{0.5L}\approx2^{0.185L}$ which implies that our time complexity exhibits a slower growth rate in relation to $L$. This finding highlights the polynomial acceleration of our approach compared to using random states, particularly in larger systems.  

Third, utilizing the optimal $M$ derived from the analysis, we sample the tensor network state to acquire the trial state by MC sampling. The MC sampling number
is from 5000 to 50000 when the size ranges from $L$ = 8 to $L$ = 20. In accordance with the Faster ground state preparation method, we conduct simulation experiments for each of the three models, with a system dimension of $L=4\times2$. The parameter $m$ was varied from 1 to 100 in our numerical test. 

\begin{figure}[!h]
 \centerline{
	\includegraphics[width=0.6\columnwidth]{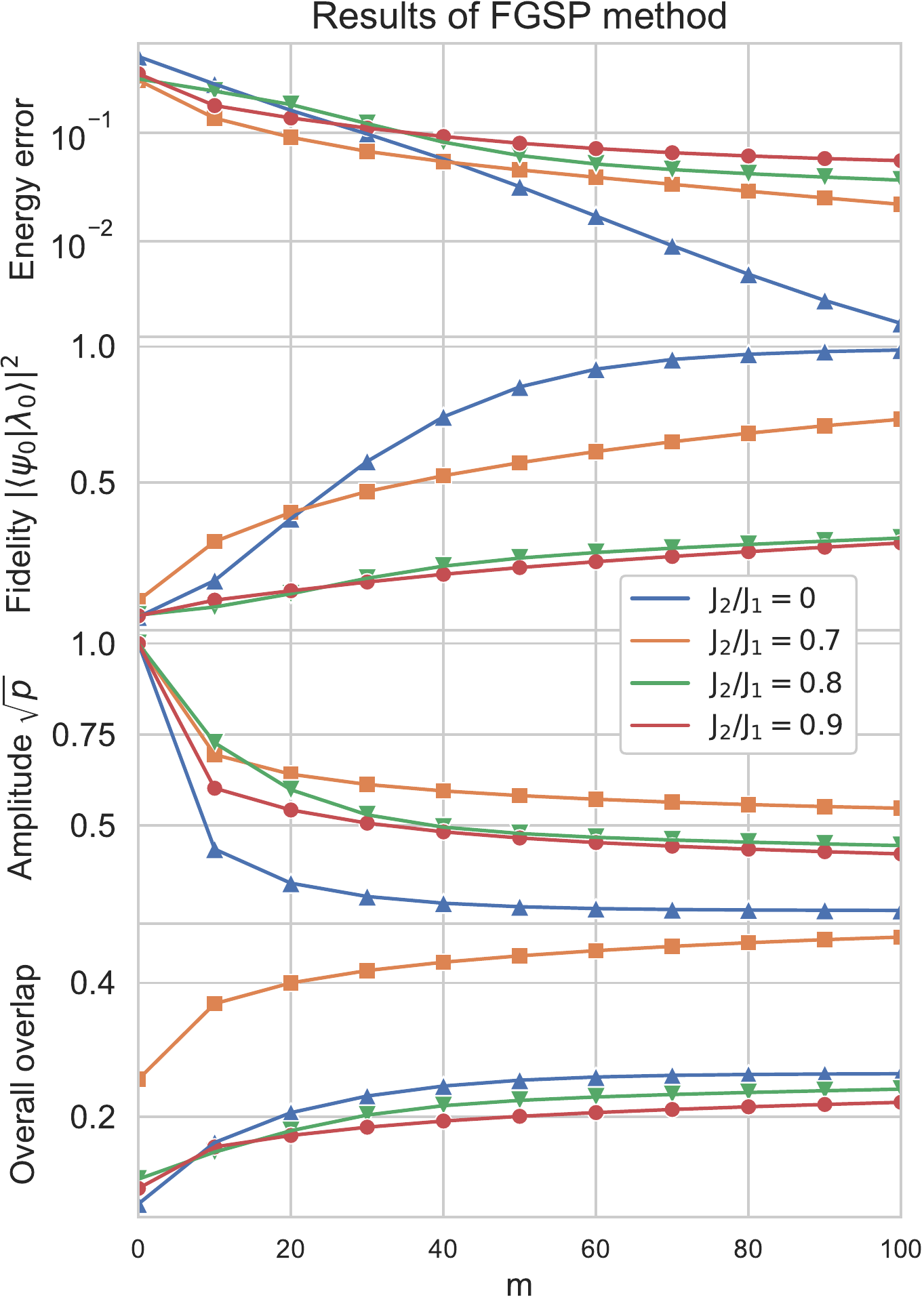}
 }
\caption{The performance of Energy error, fidelity, Amplitude, and success for $m$ ranges from 1 to 100 on the spin-$1/2$ $J_1$-$J_2$ Heisenberg model where $J_2$/$J_1=0,0.7,0.8,0.9$. The rationale behind selecting these specific values is based on the observation of a phase transition point of 0.8 depicted in Fig.~\ref{fig6}.
The energy error is the absolute value of the difference between the experimental and genuine ground state energy. Overall overlap is the product of fidelity $|\langle \psi_0|\lambda_0 \rangle|^2$ and amplitude $\sqrt{p}$, which increases as $m$ grows. We can determine the appropriate $m$ by observing convergence. }
\label{fig5}
\end{figure}

Here, we discuss the outcomes of our simulation experiments according to the state $|\varphi_2\rangle$ before amplitude amplification which is described in Appendix~\ref{appendixFGSP}.
We describe $|\varphi_2\rangle$ as
\begin{equation}
\label{equ_result}
|\varphi_2\rangle
=\sqrt{p}|0\rangle^{\otimes b}|\psi_{0}\rangle+\sqrt{1-p} |0\rangle^{\otimes b \bot }|\psi_{1}\rangle,
\end{equation}
where $\sqrt{p}$ denotes the amplitude of $|\psi_{0}\rangle$ which determines the success rate of amplitude amplification and $|\psi_{0}\rangle$ is the target state which can be rewritten as
\begin{equation}
\label{equ_psi_0}
|\psi_{0}\rangle=\langle \lambda_0|\psi_0 \rangle|\lambda_0\rangle+\sqrt{1-\langle \lambda_0|\psi_0 \rangle^2}|\lambda_0^{\bot}\rangle, 
\end{equation}
where $|\langle \psi_0|\lambda_0 \rangle|$ denotes overlap of result state $|\psi_0\rangle$ and genuine ground state.
Hence, 
\begin{equation}
\label{equ_psi_0_final}
|\varphi_2\rangle
=\sqrt{p}\langle \lambda_0|\psi_0 \rangle|0\rangle^{\otimes b}|\lambda_0\rangle+\sqrt{p-p\langle \lambda_0|\psi_0 \rangle^2}|0\rangle^{\otimes b}|\lambda_0^{\bot}\rangle+\sqrt{1-p} |0\rangle^{\otimes b \bot }|\psi_{1}\rangle.
\end{equation}
Therefore, the overall overlap of our entire quantum algorithm is absolute of $\sqrt{p}\langle \psi_0|\lambda_0 \rangle$ which consists of the amplification overlap $\sqrt{p}$ and the overlap $\langle \psi_0|\lambda_0 \rangle$ with final resultant state and genuine ground state.
The outcomes of these simulation experiments, including energy error, fidelity $\langle \psi_0|\lambda_0 \rangle^2$, amplitude $\sqrt{p}$, and overall overlap, are presented in Fig.~\ref{fig5}. 

\begin{figure}[!h]
 \centerline{
	\includegraphics[width=0.65\columnwidth]{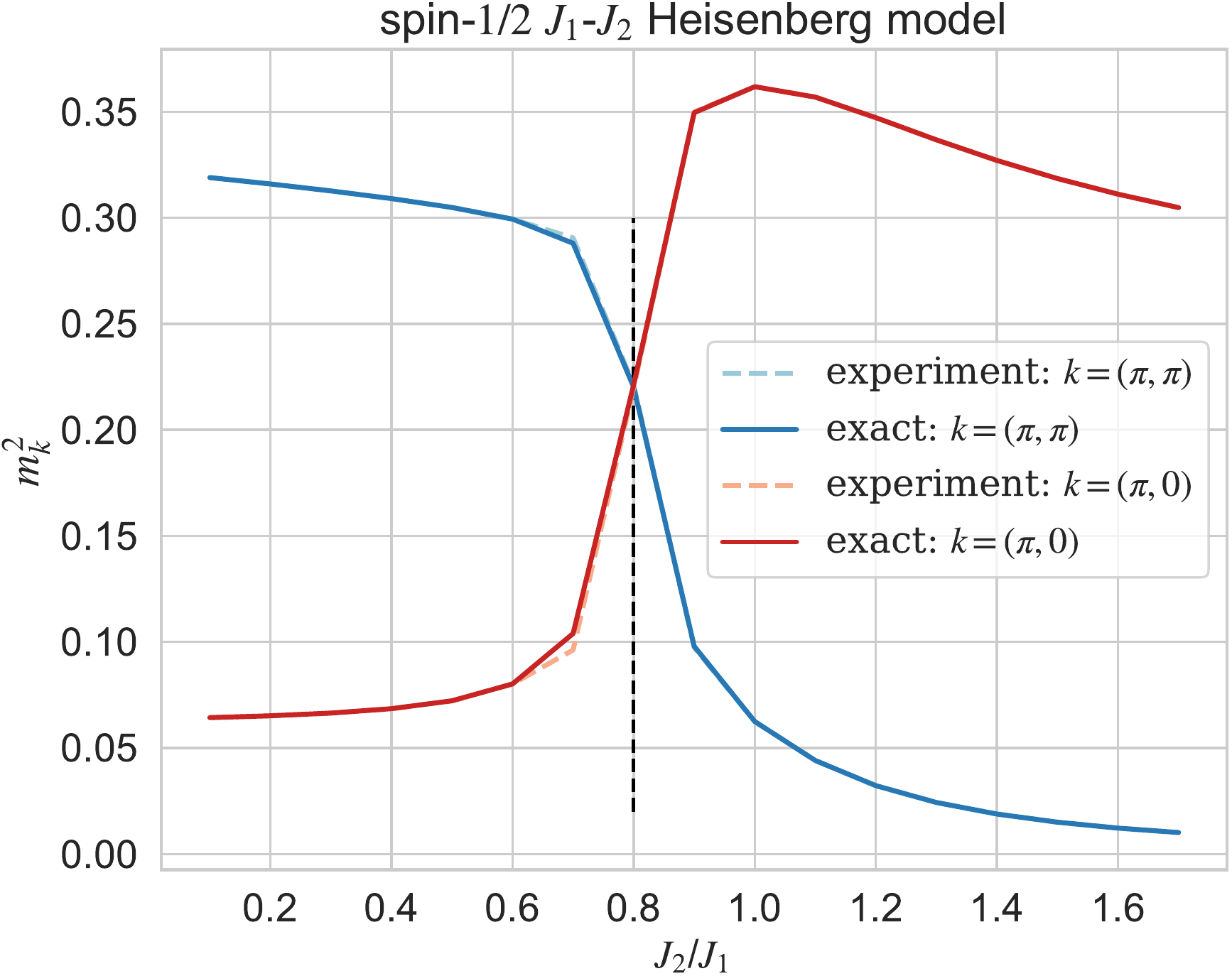}
 }
\caption{Illustration of the $m_{\bm k}^2$ at ${\bm k}=(\pi,\pi)$ and ${\bm k}=(\pi,0)$ obtained by exact ground states and experimental states where $J_2/J_1$ ranges from 0.1 to 1.8. The horizontal dashed line indicates phase transition point is 0.8 on $4\times 2$ lattice of the spin-$1/2$ $J_1$-$J_2$ Heisenberg model.}
\label{fig6}
\end{figure}

To analyze the physical property and verify the accuracy of the final result state, we calculate the magnetic structure factor $ m^2_{\bm k}=\frac{1}{L^2} \sum_{i,j}\langle \phi | S_i S_j |\phi \rangle e^{i{\bm k}({\bm r}_i - {\bm r}_j)}$ for ${\bm k}=(\pi,\pi)$ and ${\bm k}=(\pi,0)$, which are order parameters for Neel phase, and vertical stripe phase. The system undergoes a phase transition from the Neel phase to the vertical stripe phase, and the phase transition point is
$J_2/J_1$ = 0.8 on $4\times 2$ lattice. As depicted in Fig.~\ref{fig6}, it shows that our experimental results are in perfect agreement with the results of the exact ground state, indicating the accuracy of our algorithm.

In conclusion, our proposed method for ground state preparation using iTEBD and FGSP methods has demonstrated polynomial acceleration compared to using random states, by preparing a trial state that has a relatively high probability. The continuous concave nature of the algorithm's time complexity curve allows us to find the optimal M that minimizes the time complexity. These results demonstrate the polynomial acceleration of our method and show our method's potential for simulations of complex and larger systems.

\section{Conclusion}
\label{section6}
In this paper, we propose a classical assisted quantum ground state preparation method for quantum many-body systems, by combining tensor network states and Monte Carlo sampling as a heuristic method to prepare a trial state with non-trivial overlap. We employ the combination of iTEBD with TNS and FGSP~\cite{ge2019faster} on classical and quantum computers respectively. Monte Carlo sampling is a bridge to generate a trial state with a twice-approximated process. We demonstrate the effectiveness of our approach through numerical tests in spin-$1/2$ $J_1$-$J_2$ Heisenberg model. The numerical test demonstrates a polynomial acceleration in the scaling of overlap between the trial state and genuine ground state compared to random trial states, reducing from $2^{0.5L}$ to $2^{0.5\delta L}$ and $0<\delta<1$. Notably, $\delta$ is only 0.37 in the case of $J_2/J_1=0$, which implies that our time complexity exhibits a slower growth rate in relation to $L$, resulting in a more pronounced difference as $L$ increases. Moreover, the states obtained by our method are highly accurate and exhibit a significant phase transition. Furthermore, our method has the potential to be extended to other tensor network methods and quantum algorithms. For instance, they can be replaced by variational methods~\cite{Low_2010,reiner2019finding}, the Density Matrix Renormalization Group (DMRG)~\cite{white1992density,danshita2009bose,vidal2004efficient} or near-optimal quantum ground state preparation~\cite{lin2020near}.

While our method shows significant results, there is still room for improvement.
Firstly, determining the optimal value of $M$ relies on graphing the method's time complexity, which could be simplified if an analytical solution were derived by directly taking partial derivatives. 
Secondly, although we have demonstrated the feasibility of our method through numerical tests, the efficiency of the three specifications of the method mentioned in Fig.~\ref{trial_state} are currently conjectures and lack theoretical proof. In future research, it would be valuable to analyze the nature of the ground state and potentially obtain a more accurate model analysis.

Overall, our work provides a method for effectively preparing a trial state with non-trivial overlap that is used in certain quantum ground state preparation methods in quantum many-body systems. We hope that our work will inspire further research in this area and be applied to larger-scale models and real quantum computers.

\section{Acknowledgement}
This work was supported by the National Natural Science Foundation of China (No. 12104433).

\newpage
\bibliographystyle{unsrt}

\onecolumn\newpage
\appendix

\section{Derivation of Complexity}
\label{derivation}

\begin{figure}[!h]
 \centerline{
	\includegraphics[width=0.55\columnwidth]{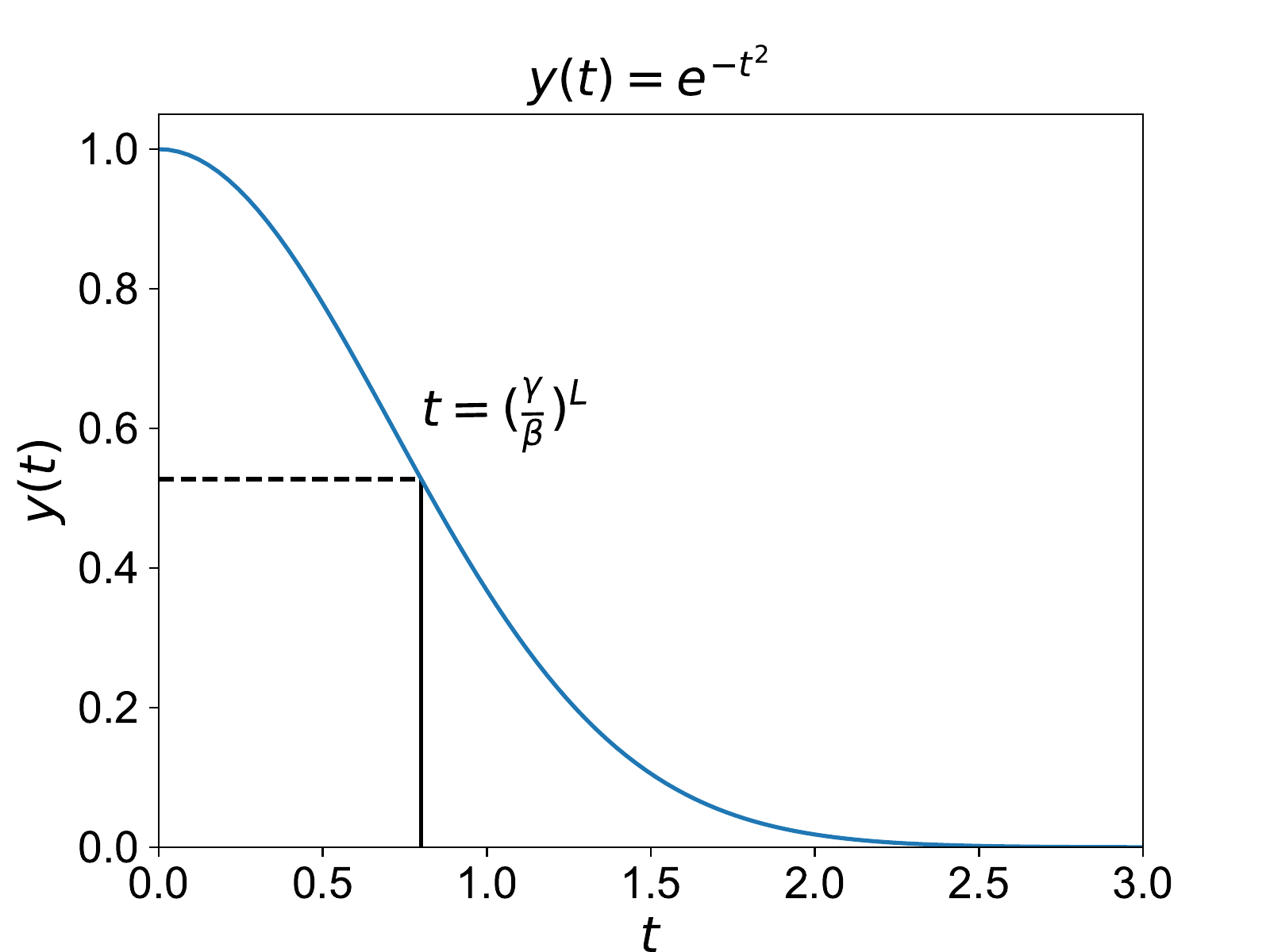}
 }
\caption{Variable upper limit integral of $e^{-t^2}$ for $(\gamma/\beta)^L$. The solid black line denotes $t=(\gamma/\beta)^L$ and the black dashed line represents the corresponding y-axis coordinates. We assume $0<(\gamma/\beta)^L<1$ since normally $\gamma<\beta$ in our numerical test.}
\label{integral}
\end{figure}

As described in Eq.~(\ref{Tfinal}), we focus on exponential growth and ignore polynomials.
Our analysis centers around the complexity of $\operatorname{\bm erf}((\gamma/\beta)^L)$ in the denominator.
We visualize the variable upper limit integral in Fig.~\ref{integral}.
The integration region spans from 0 to $(\gamma/\beta)^L$ beneath the function $y=e^{-t^2}$.
We let $x=(\gamma/\beta)^L$ and approximate this region as the area of the rectangle framed by the black line in the figure.
Then, the integral value can be approximated as $xe^{-x^2}$. To further evaluate this, we perform a Taylor expansion on $e^{-x^2}$, yielding 
\begin{equation}
x(1-x^2+\frac{x^4}{2}+{\mathcal{O}}(x^6)),
\end{equation}
where $0<x<1$ normally in our numerical test. Thus at large $L$, we seem to ignore values above the second order. 
\begin{equation}
T_{\operatorname{erf}}=\widetilde{\mathcal{O}}((\gamma/\beta)^L),
\end{equation}
which can be substituted into Eq.~(\ref{Tfinal}), obtain
\begin{equation}
T=\widetilde{\mathcal{O}}(\gamma^{L/2}(\gamma^{L/2}+\beta^{L/2})).
\end{equation}
Furthermore, since $\gamma<\beta$, we can approximate
\begin{equation}
T=\widetilde{\mathcal{O}}((\gamma\beta)^{L/2}).
\end{equation}

\section{Additional Numerical Tests}
\label{appendix1}
\begin{figure}[!h]
 \centerline{
	\includegraphics[width=0.55\columnwidth]{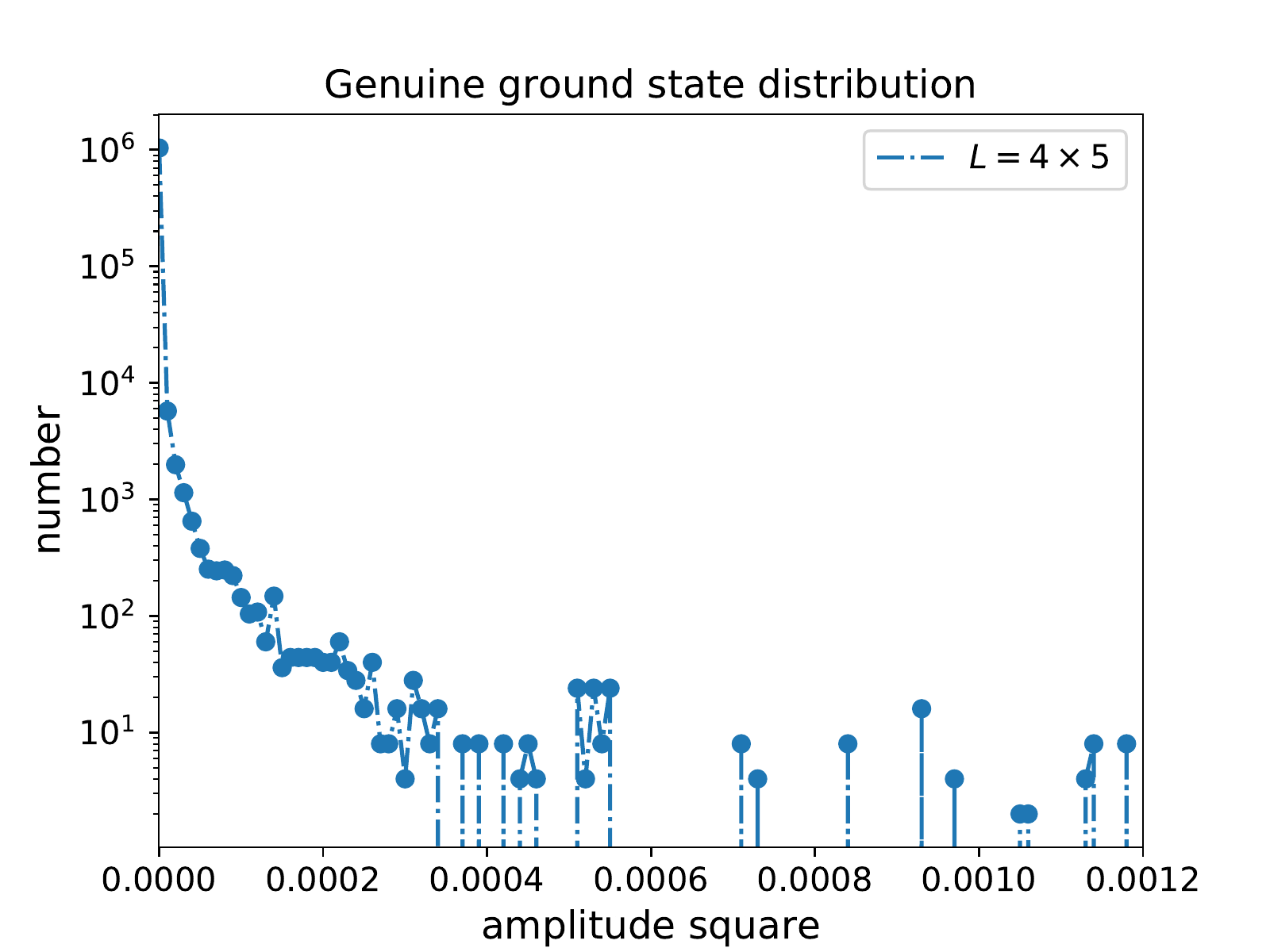}
 }
\caption{The distribution of amplitude square for the exact ground state in the scale of $L=4\times5$ in spin-$1/2$ $J_1$-$J_2$ Heisenberg model where $J_2/J_1=0$.}
\label{fig9}
\end{figure}

In this section, we provide some of the experimental data to complement our proposed algorithm. First, we assume there exists a Gaussian error relationship between $M$ and $f$. The distribution of the exact ground state for the spin-$1/2$ $J_1$-$J_2$ Heisenberg model is Normal distribution in a system size of $L=4\times5$, as shown in Fig.~\ref{fig9}.

\begin{figure}[!h]
 \centerline{
	\includegraphics[width=0.6\columnwidth]{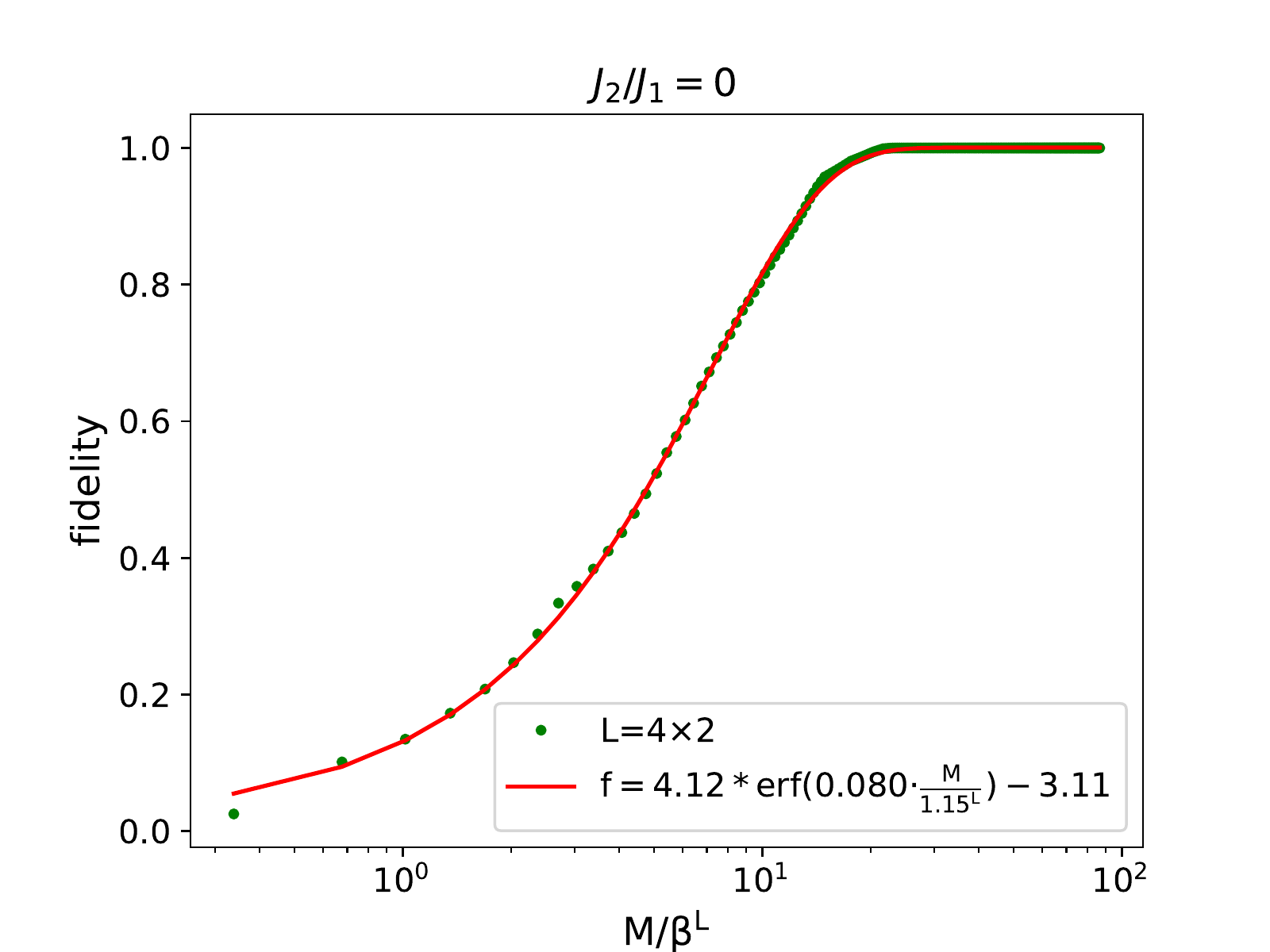}
 }
\caption{The distribution of fidelity $f_1$ between PEPS and genuine ground state and major components $M$ is analyzed in in spin-$1/2$ $J_1$-$J_2$ Heisenberg model where $J_2/J_1=0$ and $L=4\times2$. The horizontal coordinates represent the interval of the probability squared, divided by 0.00001. The vertical coordinates indicate the number of individuals in the x-axis interval.}
\label{fig10}
\end{figure}

For instance, for a system size of $L=4\times2$, the fitting curve has high overlap with original data shown in Fig.~\ref{fig10}. Next, in order to reconcile these contrasting time complexities, a parameter $\delta$ is introduced to achieve a harmonized integration. Then, we plot the time complexity by Eq.~(\ref{equ5}) and search the optimal $M$. 

\begin{figure}[!htp]
 \centerline{
	\includegraphics[width=0.6\columnwidth]{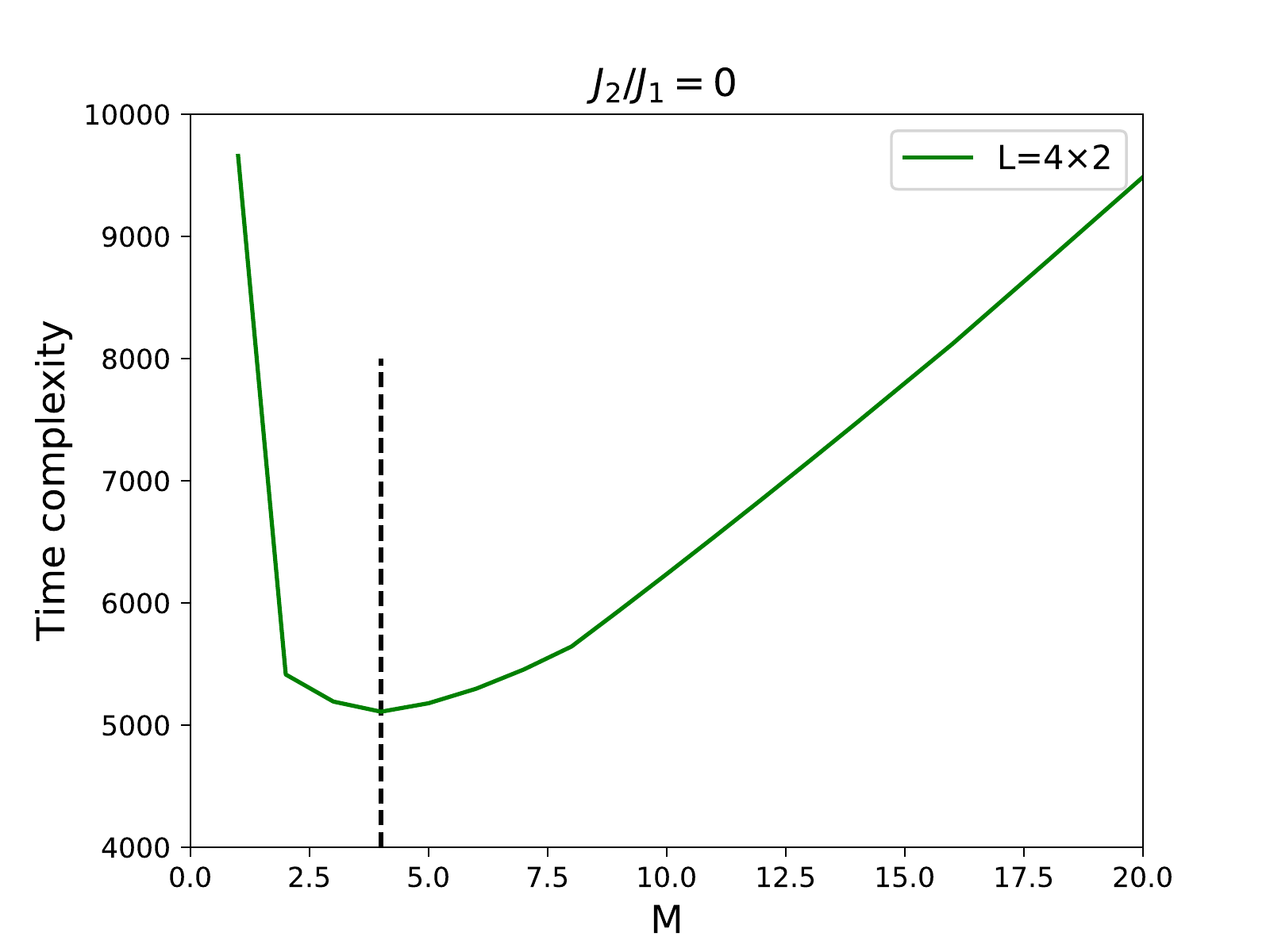}
 }
\caption{The time complexity for various $M$ on the Heisenberg model with a size of $L=4\times2$. The black dashed line represents the optimal point $M=4$.}
\label{fig11}
\end{figure}

\section{Faster Ground State Preparation}
\label{appendixFGSP}

Faster Ground State Preparation is a method to filter the trial state by employing LCU lemma and a linear combination of time-evolutions of different time lengths, which achieves the complexity up to logarithmic factors~\cite{ge2019faster}. 

A condition for this algorithm is that the spectrum of Hamiltonian is contained in [0, 1]. 
Assuming knowledge of the lowest eigenvalue $\lambda_0$ and maximum eigenvalue $\lambda_{m}$ of $H$, it is straightforward to transform into $\widetilde{H} = (H-\lambda_0 I)/ \lambda_{m}$. 
Then, prepare a state $\epsilon$-close to $|\phi_0\rangle$ by (approximately) projecting $|\varphi_1\rangle$ onto ground state of $\widetilde{H}$.
Given that $\Lambda$ represents the ``base cost'' of the simulation, $\Delta$ denotes the established minimum value for the spectral gap of $H$, and $\Phi$ is responsible for preparing a trial state. Let $\chi$ be a known lower bound on overlap $|\langle\phi_{0}|\varphi_{1}\rangle|$ and assume that $\chi = e^{-\mathcal{O}(\log N)}$ throughout this paper. 

First, approximate ${\cos}^{m}H$ as a linear combination of terms of the form $e^{-iHt_k}$.
This was proved by assuming $m=2n$ for simplicity and LHS represents the probability of observing more than $n + m_0$ heads when flipping $2n$ coins in~\cite{ge2019faster}. Thus,
\begin{equation}
\cos^{2n}H=\sum_{k=-m_0}^{m_0} \alpha_k e^{-2iHk}+\mathcal{O}(\chi \epsilon),
\end{equation}
where
\begin{equation}
\alpha_k:=2^{-2n}
\begin{pmatrix}
 2n\\n+k
\end{pmatrix}
\end{equation}
and
\begin{equation}
m_0=\Theta(\frac{1}{\Delta}\log^{3/2}\frac{1}{\chi\epsilon})
\end{equation}
Then, we employ the LCU lemma to implement $e^{-2iHk}$: Let B be a circuit on $b:=\lceil \log_2(2m_0+1) \rceil$ qubits that maps $|0\rangle^{\otimes b}$ to 
\begin{equation}
B|0\rangle^{\otimes b}:=\frac{1}{\sqrt{\alpha_s}}\sum_{k=-m_0}^{m_0} \sqrt{\alpha_k}|k\rangle,
\end{equation}
where $\alpha_s=\sum_{k=-m_0}^{m_0}\alpha_k$ and let U be the controlled Hamiltonian simulation U$|k\rangle|\varphi_1\rangle=|k\rangle e^{-2iHk}|\varphi_1\rangle$. Then,
\begin{equation}
|\varphi_2\rangle = (B^\dagger \otimes \mathbbm{1})U(B \otimes \mathbbm{1}) |\varphi_1\rangle =\frac{1}{\alpha_s}|0\rangle^{\otimes b}\sum_{k=-m_0}^{m_0} \alpha_k e^{-2iHk} |\varphi_1\rangle +|R\rangle,
\end{equation}
Finally, we apply amplitude amplification. By separating the target and non-target states, we can rewrite 
\begin{equation}
\label{equ3}     
|\varphi_2\rangle=\sqrt{p}|0\rangle^{\otimes b}|\psi_{0}\rangle+\sqrt{1-p} |0\rangle^{\otimes b \bot }|\psi_{1}\rangle,
\end{equation}
where $|\psi_{0}\rangle$ is the target state and $p$ denotes the success rate. 

\begin{figure}[!h] 
  \centerline{
\Qcircuit @C=0.8em @R=1.0em {
&&-S_{\chi} & &A^{-1}  &  &S_{0} &  &A\\
\lstick{|1\rangle}& \qw &\gate{Z}  &\qw &\qw &\qw  &\gate{Z} &\qw &\qw &\qw & \qw \\
\lstick{|0\rangle}& \qw&\ctrlo{-1}&\qw &\multigate{1}{A^{-1}} &\qw&\ctrlo{-1}&\qw&\multigate{1}{A}  & \qw&\qw \\
& \qw& \qw &\qw &\ghost{A^{-1}} &\qw&\ctrlo{-1} &\qw&\ghost{A}  & \qw&\qw 
}
}
\caption{Quantum circuits of $Q=-AS_{0}A^{-1}S_{\chi}$. According to the expression, the quantum circuit involves four parts from left to right. Each part corresponds to the topmost marker. The input state consists of an auxiliary qubit $|1\rangle$ and $|\psi\rangle$. The gate $A$ is the same as $A$ in Fig.~\ref{fig2}.}
\label{fig3}
\end{figure}
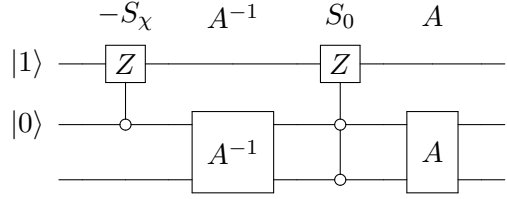

\begin{figure}[!h]
 \centerline{
	\includegraphics[width=0.7\columnwidth]{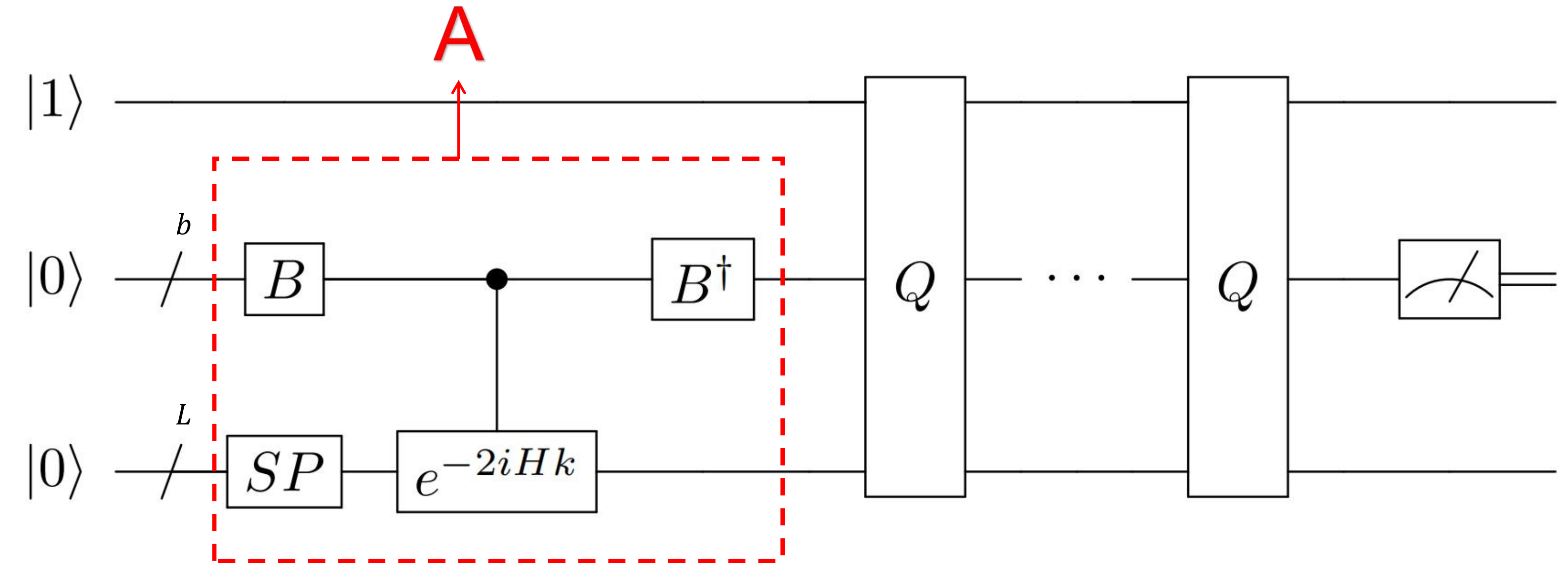}
 }
\caption{Quantum circuit of the quantum ground state preparation algorithm. The whole circuit involves two parts: linear combination and amplitude amplification which is consisting of several $Q$ gates. The top is an auxiliary bit for amplitude amplification. Gate $SP$ denotes state preparation, and it maps $|0\rangle^{\otimes L}$ to trial state $|\varphi_{1}\rangle$ which is sampled from TNS. }
\label{fig2}
\end{figure}

The expression of Q has been elaborated that $Q=-AS_{0}A^{-1}S_{\chi}$~\cite{brassard2002quantum}. Here, the unitary transformation $A$ is the same as $A$ in Fig.~\ref{fig2}. The function of $S_{0}$ is to flip the amplitude of $|\psi\rangle$ when it is all 0 and $-S_{\chi}$ is to flip the amplitude of $|\psi^{'}_{0}\rangle$ when $|q\rangle$ is 0. 
This flip can be achieved by applying the Z gate to an auxiliary qubit $|1\rangle$. Therefore, the circuit of $Q$ is constructed as shown in Fig.~\ref{fig3}. And the total quantum circuit is shown in Fig.~\ref{fig2}.

However, the number of QAA executions $j$ depends on $|\psi_{0}|$ which is the measured probability that $|\varphi_2\rangle$ yields our target state. When the amplitude of $|\psi_{0}\rangle$ is unknown, Quantum Amplitude Estimation (QAE) is an effective technique that aims to estimate the amplitude of a given quantum superposition state.

\end{document}